\documentclass{article}
\usepackage{arxiv}
\usepackage{tikz}
\usepackage{graphicx,color}
\usepackage[fleqn]{amsmath}
\usepackage{amsfonts,verbatim}
\usepackage{amssymb}
\usepackage{epsfig}
\usepackage{bm}
\usepackage{subfigure}
\usepackage{array}
\usepackage{cleveref}

\usepackage{algorithm} 
\usepackage[noend]{algpseudocode}
\makeatletter
\def\BState{\State\hskip-\ALG@thistlm}
\usepackage{changepage}

\begin{document}

\title{Bayesian uncertainty quantification for micro-swimmers with fully resolved hydrodynamics}

\author{
  Karen Larson \\ 
  Division of Applied Mathematics\\ 
  Brown University \\
  Providence, RI 02912 USA\\
  \texttt{karen\_larson@brown.edu} \\
   \And
 Sarah Olson \\
  Department of Mathematical Sciences\\
  Worcester Polytechnic Institute\\
  Worcester, MA 01609 USA \\
  \texttt{sdolson@wpi.edu} \\
  \And
  Anastasios Matzavinos \\ 
  Division of Applied Mathematics\\ 
  Brown University \\
  Providence, RI 02912 USA \\
  \texttt{matzavinos@brown.edu} \\
}
\date{\today}
\maketitle 

\begin{abstract}
Due to the computational complexity of micro-swimmer models with fully resolved hydrodynamics, parameter estimation has been prohibitively expensive. Here, we describe a Bayesian uncertainty quantification framework that is highly parallelizable, making parameter estimation for complex forward models tractable. Using noisy \textit{in silico} data for swimmers, we demonstrate the methodology's robustness in estimating the fluid and elastic swimmer parameters. Our proposed methodology allows for analysis of real data and demonstrates potential for parameter estimation for various types of micro-swimmers. Better understanding the movement of elastic micro-structures in a viscous fluid could aid in developing artificial micro-swimmers for bio-medical applications as well as gain a fundamental understanding of the range of parameters that allow for certain motility patterns.
\end{abstract}


\section{Introduction}
Mathematical models to investigate the swimming speeds and efficiency of micro-swimmers date back to the foundational work of Taylor \cite{Gaffney11,Lauga09,Taylor51,Taylor52}. This pioneering work focused on studying a swimmer represented as either a planar sheet in 2-dimensions (2D) or a waving cylindrical tail in 3-dimensions (3D), both of infinite extent and with a prescribed beat form. State of the art now involves highly computational 2D or 3D models of swimmers with a finite length flagellum (tail) and accurate geometries where the beat form is either prescribed or an emergent property of the coupled system where the fluid dynamics are fully resolved \cite{Carichino18,Dillon06,Elgeti10,Elgeti15,Huang18,Ishimoto18,Schoeller18,Smith09Surf,Toshihiro19,Yang08}. These models are able to to incorporate material properties of the actual flagellum and the fluid stress may be linear or nonlinear with respect to the strain \cite{Ishimoto18b,Olson11,Teran10,Thomases14}.

Currently, there is great interest in understanding the coupling between fluids and the swimming structure. In the case of sperm motility, understanding motility patterns could lead to the development of treatments to either aid or hinder sperm progression \cite{Mortimer00}. In experiments, CASA (computer aided sperm analysis) is a widely used software that is able to track the head or cell body of sperm to track swimming velocities and can also be used to determine beat frequency of the flagellum (tail). However, these data are often noisy and one might want to understand the fluid properties or swimmer properties (geometry or constitutive parameters) that would allow for these types of trajectories. 
Previous estimation of constitutive parameters has usually been done without fluid-structure interaction models or without fully resolved hydrodynamics. Parameter regimes of interest are generally inferred from experimental data and/or solving a simplified 1-d type model  \cite{Dominic09,Gadelha13,Riedel07}. This of course is very restrictive.

In addition, an active area is in the development of artificial micro-swimmers for applications such as drug delivery and microsurgery \cite{Gao14,Nelson10,Sanders09,Tierno08}. In the case of artificial micro-swimmers, most are bio- inspired and have flagella. The swimming is achieved through physical mechanisms, and could be driven by a magnetic field, electric field, or a chemical reaction. In the context of artificial micro-swimmers, one might want to determine parameters corresponding to an optimal swimming gate for a particular application. On the other hand, one might want to understand whether the variation in swimmer geometries and material parameters are optimal in different fluid environments. 

Little prior work has been completed in terms of parameter estimation or uncertainty quantification for micro-swimmers. Plouraboue et al. \cite{Plouraboue17} used a simplified bead model to represent a swimmer with linear elasticity. 
In this simple model,  identifiability conditions determined that parameter identification was possible in the presence of noise.
 Recently, Tsang et al.\cite{Tsang19} explored the idea of self-learning to optimize locomotion of synthetic micro-swimmers in the different fluid environments that might be encountered in the body for drug delivery applications. These artificial micro-swimmers were self-learning and adaptive. 
 
In this work, we utilize a Bayesian framework for uncertainty quantification to understand the emergent trajectories and swimming speeds of micro-swimmers. We assume the micro-swimmers are immersed in a fluid with a sparse network of fibers, which is captured via a flow dependent term with a resistance parameter, and the hydrodynamics are fully resolved utilizing the method of regularized fundamental solutions \cite{Cortez10,Leiderman16,Ho19}. In the swimmer model, we include non-linearities in the  elasticity portion since the elastic energy depends on curvature-- a nonlinear function of shape. In this context, Bayesian estimation methods allow us to leverage our prior knowledge of micro-swimmers and the viscous medium to inform the search of the parameter space. We examine the validity of using the parallel transitional Markov chain Monte Carlo (TMCMC) algorithm for micro-swimmer questions. This is done by using the high performance computing framework $\Pi$4U \cite{Hadjidoukas15, Bowman18, Larson19} to estimate parameters for the fluid resistance and material parameters of the flagella using noisy \textit{in silico} observations generated from the biological model, which allows us to garner correlations between parameters and gain insights into the underlying biological mechanisms.

\section{Micro-swimmer Model}\label{sec:swimmermodel}
\subsection{Fluid Equations}\label{fluid}
 Micro-swimmers navigate in the regime where viscous forces dominate and acceleration is negligible \cite{Lauga09,Elgeti15}. A common approach is to model the fluid that the swimmer is immersed in via the steady Stokes equations. However, many micro-swimmers encounter non-homogeneous fluids where there is a protein network or other obstacles. For example, mammalian spermatozoa navigate in vaginal and cervical fluids with a network of mucin fibers \cite{Rutllant05,Saltzman94} and bacteria navigate through a mucus layer or in biofilms, both consisting of a network of polymeric substances \cite{Flemming10,Miradbagheri16}. The appropriate governing equation for the fluid will depend on several factors relating to the volume fraction, material properties, and movement of the immersed fibers or proteins.  

In the case of a stationary and sparse network of fibers or obstacles in the fluid, the Brinkman equation can be used \cite{Auriault09,Brinkman47,Durlofsky87,Howells74,Spielman68}. This is a fluid model that does not model the exact architecture and location of the fibers, but captures the homogenized fluid flow, and has been successfully used to model different types of micro-swimmers \cite{Fu10,Leshansky09,Morandotti12,Nganguia18,Ho19}. The incompressible Brinkman equation is
\begin{equation}
\mu_f\Delta \bm{v^*} - \frac{\mu_f}{{K_D}} \bm{v^*} = \nabla \mathrm{p}^* - \bm{{F}^*},\hspace{.3cm}
\nabla \cdot \bm{v^*} = 0
\end{equation}
where $\bm{v}^*$ is the fluid flow, $\mathrm{p}^*$ is the pressure, $\mu_f$ is the viscosity of the fluid, and the fibers or particles are represented via the darcy permeability $K_D$ (units of length squared). Here, $\bm{{F}^*}$ is a force density that a micro-swimmer exerts on the surrounding fluid \cite{Ho16,Ho19,Leiderman16,Olson15}. We use an immersed boundary approach where the swimmer is assumed to be a neutrally buoyant structure immersed in the incompressible fluid \cite{Peskin02}. We note that for micro-swimmers, this governing equation is only valid when the fibers are sparse enough such that there is room for the swimmer to navigate without pushing on any of the fibers. Spielman \cite{Spielman68} derived a relationship between the darcy permeability and the volume fraction of fibers and we have previously shown that a volume fraction of less than 2\% fibers is valid and in a biologically relevant range for micro-swimmers such as sperm \cite{Ho19,Ho16,Leiderman16}.  

Defining $L$ as the characteristic length scale (length of the micro-swimmer) and $\alpha= L/\sqrt{K_D}$ as the resistance parameter, we
arrive at the nondimensional Brinkman equation,
\begin{equation}
\Delta \bm{v} - \alpha^2 \bm{v} = \nabla \mathrm{p} - \bm{{F}},\hspace{0.3cm}\nabla \cdot \bm{v} = 0. \label{BrFlow}
\end{equation}
This equation governs the fluid flow $\bm{v}$ at any point $\bm{x}$ in the domain. Since the force density corresponding to the micro-swimmers will be a singular force layer, we write $\bm{{F}}(\bm{x})={\bm{\mathcal{F}}}\delta(\bm{x}-\bm{X})$ where the micro-swimmer location(s) are given by $\bm{X}$ and $\delta$ is the delta distribution. Since the Brinkman equation is linear, we can utilize fundamental solutions to solve for the resulting flow. However, swimmers represented as curves will lead to singular integrals. To resolve this issue, we can regularize the forces using a mollifier $\phi_{\varepsilon}(\bm{x}-\bm{X})$ and then solve for the resulting regularized fundamental solution for the Brinkman equation, known as the regularized Brinkmanlet \cite{Cortez10,Leiderman16}. In this case, $\bm{{F}}$ in (\ref{BrFlow}) is replaced with $\bm{\mathcal{F}}\phi_{\varepsilon}(\bm{x}-\bm{X})$, where $\varepsilon$ is the regularization parameter corresponding to the region where most of the force is spread to the fluid. In the case of a micro-swimmer, we can choose this width to correspond to the radius of the flagellum (tail). 

We will study micro-swimmers such as sperm in a 2-dimensional, infinite fluid. Each swimmer is represented as a discretized set of points, $\bm{X}_k$ for $k=1,\ldots,N$. We choose the following radially symmetric mollifier
\begin{equation}
\phi_{\varepsilon}(r)=\frac{3\delta^3}{2\pi (r^2 + \delta^2)^{5/2}},
\label{eqnblob}
\end{equation}
that satisfies  $\int_0^{\infty}r\phi_{\delta} dr=\frac{1}{2\pi}$ where $r=|\bm{x}-\bm{X}_i|$ for a point $\bm{X}_i$ on the swimmer (utilizing the standard Euclidean norm). The solution approach involves taking the divergence of the regularized version of (\ref{BrFlow}) to solve for the pressure, which is then plugged back into (\ref{BrFlow}) to solve for the resulting Brinkman flow. Since the Brinkman equation is linear, we have a solution that is a superposition of all of the point forces. As described in \cite{Leiderman16}, the resulting fluid flow and pressure at a point $\bm{x}$ due to $\mathcal{N}$ regularized point forces on the structure of the swimmer is given as
\begin{subequations}
\begin{align}
\mathrm{p}(\bm{x})&=\sum_{i=1}^\mathcal{N}\bm{\mathcal{F}_i}\nabla G_{\varepsilon}(r),\label{BrP}\\
\bm{v}(\bm{x})&=\sum_{i=1}^\mathcal{N}\left[-\bm{\mathcal{F}_i}B_{\varepsilon}''(r) + (\bm{\mathcal{F}}_i \cdot (\bm{x}-\bm{X}_i))(\bm{x}-\bm{X}_i)\frac{rB_{\delta}''(r) - B_{\delta}'(r)}{r^3}\right], \label{BrV}
\end{align}
\end{subequations}
where $\phi_{\varepsilon}=\nabla^2G_{\varepsilon}$ and $B_{\varepsilon}$ is implicitly defined as $(\Delta-\alpha^2)B_{\varepsilon}=G_{\varepsilon}$. Details of the numerical method are given in Leiderman et al.\cite{Leiderman16,Leiderman17}. For the swimmer, at each point in time, it will have a given configuration $\bm{X}$ that will lead to forces $\bm{\mathcal{F}}$ along the structure (details in the Appendix). Thus, at each moment in time, we are solving for the instantaneous fluid flow. To march forward in time, we assume that the swimmer is moving with the resulting fluid flow, i.e. $d\bm{X}/dt=\bm{v}(\bm{X})$. 
In the case of $\mathcal{M}_S$ swimmers each with $\mathcal{N}_T$ points, the summation in (\ref{BrP})--(\ref{BrV}) is then taken over the $\mathcal{M}_S\mathcal{N}_T$ points. This becomes computationally intensive when solving for a large number of time steps or for a large number of micro-swimmers since there are approximately $\mathcal{O}(\mathcal{M}_S\mathcal{N}_T)$ calculations at each time step.

The micro-swimmer we model is a simplified representation of a sperm with a head (cell body) attached to a flagellum (tail).  Since we are at zero-Reynolds number, this will correspond to force and torque-free swimming \cite{Lauga09}. (Note that in the case when the sum of forces in the system is non-zero, extra terms must be accounted for in (\ref{BrV}) \cite{Ahmadi17,Leiderman17}.) The active force generation along the elastic flagellum propels the swimmer forward whereas the head is a passive and fairly rigid structure (containing the genetic material) \cite{Gaffney11}. The flagellum is discretized into $\mathcal{N}_F$ points ($\bm{X}_F^j$) and the head  is discretized into $\mathcal{N}_H$ points ($\bm{X}_H^j$), for a total of $\mathcal{N}_T=\mathcal{N}_F+\mathcal{N}_H$ point forces on each micro-swimmer ($j=1,\ldots,\mathcal{M}_S$). We detail the specifics of the force models in the Appendix. Briefly, using a model similar to Fauci and McDonald \cite{Fauci95}, we derive an energy $E(\bm{X}^j)$ for each swimmer $j$, where a variational derivative of this energy results in the forces,  
\begin{align}
    \bm{\mathcal{F}}^j&=-\frac{\partial E^j}{\partial \bm{X}^j} \notag \\
    &=-\frac{\partial}{\partial \bm{X}^j}\left(E^j_{F,bend}(\bm{X}_F^j)+E^j_{F,tens}(\bm{X}_F^j)+E^j_{H,bend}(\bm{X}_H^j)+E^j_{H,tens}(\bm{X}_H^j)+E^j_{N}(\bm{X}_F^j,\bm{X}_H^j)\right),\label{Fenergy}
\end{align}
and this is the force density that is put into (\ref{BrFlow}) as a regularized force, $\bm{F}=\bm{\mathcal{F}}\phi_{\varepsilon}$. The total energy $E(\bm{X})$, will be a sum of different energy components corresponding to a bending and tensile energy in the flagellum ($E_{F,bend}$ and $E_{F,tens}$) and head ($E_{H,bend}$ and $E_{H,tens}$), as well as an energy component corresponding to the connection of the head and flagellum in the neck region ($E_{N}$). The bending energy will drive the dynamics and resulting motion of the flagellum and is detailed in the Appendix.

\section{Bayesian Methods}\label{sec:bayesianmethods}
It is reasonable to assume that observed data for micro-swimmers will not exactly match with any model, as models are not perfect and observation data are noisy. Bayesian uncertainty quantification (UQ) allows us to approach this problem by assuming that parameters are random variables with unknown distributions and leverages prior information, knowledge, and experience to inform searches about distributions of unknown parameters.
\subsection{Parameter Estimation}\label{param}
Here, the parameters of interest $\bm{\theta}$ are the inputs for a micro-swimmer model $M$ that predicts quantities of interest $\bm{g}(\bm{\theta}|M) \in \mathbb{R}^m$, e.g. the velocity of a single micro-swimmer in the fluid or the distance between two micro-swimmers. As models cannot exactly represent physical, observed quantities $\bm{D}$ due to various errors (e.g. modeling, computational, and measurement), we need an explicit relationship between the model outputs and the noisy observed data. One possible relationship is that the observed data $\bm{D}$ are generated according to the model prediction equation:
\begin{equation}
\bm{D} = \bm{g}(\bm{\theta}|M) + \bm{e},
\label{eq:model_err}
\end{equation}
where $\bm{e}$ is the prediction error and $\bm{g}(\bm{\theta}|M)$ are the model predictions for a given set of parameters $\bm{\theta} \in \mathbb{R}^n$. 

The posterior distribution of the parameters given the observations is given by Bayes' formula as: 
\begin{equation}
p(\bm{\theta}|\bm{D},M) = \dfrac{p(\bm{D}|\bm{\theta},M)\pi(\bm{\theta}|M)}{\rho(\bm{D}|M)},
\label{eq:bayes}
\end{equation} 
where $p(\bm{D}|\bm{\theta},M)$ is the likelihood that the observations came from our model $M$ with parameters $\bm{\theta}$ as inputs, $\pi(\bm{\theta}|M)$ is the prior distribution on the parameters, and $\rho(\bm{D}|M)$ is the evidence of the model class, given by the multi-dimensional integral
\begin{equation*}
\rho(\bm{D}|M) = \int_{\mathbb{R}^n}p(\bm{D}|\bm{\theta},M)\pi(\bm{\theta}|M)d\bm{\theta}.
\end{equation*}
When $M$ is one particular model for a system of interest in a family of parameterized models, $\rho(\bm{D}|M)$ serves as a measure of fit for how well the model matches the observed data \cite{Larson19, Bowman18,Beck04}. For the parameter estimation problem that we focus on here, we only need to compute the likelihood $p(\bm{D}|\bm{\theta},M)$ and the prior $\pi(\bm{\theta}|M)$, as $\rho(\bm{D}|M)$ serves as a normalization constant since it is independent of $\bm{\theta}$.

In order to calculate the likelihood $p(\bm{D}|\bm{\theta},M)$ in \eqref{eq:bayes}, we need to postulate a form for the error term $\bm{e}$. Here, we assume that the error is normally distributed with zero mean and covariance matrix $\bm{\Sigma}$. In addition, we assume that the errors at different times are uncorrelated, so that the covariance matrix takes on the form $\bm{\Sigma} = \sigma \bm{I}$ where $\bm{I}$ is the $m \times m$ identity matrix.

Since $M$ is a deterministic model, it follows that $\bm{D}$ is also normally distributed, and the likelihood takes on the form \cite{Larson19, Bowman18, Beck04, Vanik00}
\begin{equation}
p(\bm{D}|\bm{\theta},M) = \dfrac{|\bm{\Sigma}(\bm{\theta})|^{-1/2}}{(2\pi)^{m/2}}\exp\left[-\frac{1}{2}J(\bm{\theta},\bm{D}|M)\right],
\label{eq:likelihood}
\end{equation}
where 
\begin{equation}
J(\bm{\theta}, \bm{D}|M) = [\bm{D} - \bm{g}(\bm{\theta}|M)]^T\bm{\Sigma}^{-1}(\bm{\theta}) [\bm{D} - \bm{g}(\bm{\theta}|M)]
\label{eq:fitness}
\end{equation}
is the weighted measure of fit between the data and the model predictions, $| \cdot |$ denotes determinant, and the parameter set $\bm{\theta}$ is augmented to include the parameters involved in the structure of the covariance matrix $\bm{\Sigma}$ (here, the noise level $\sigma)$.

The main computational bottleneck in parameter estimation is the computation of the complex forward model $\bm{g}(\bm{\theta}|M)$. The $\Pi$4U framework  \cite{Hadjidoukas15} used has the advantages of using a massively parallelizable sampling algorithm, transitional Markov chain Monte Carlo \cite{Ching07}, and of having an efficient parallel structure for task sharing.

\subsection{Transitional Markov Chain Monte Carlo}\label{tmcmc}
The TMCMC algorithm slowly transitions from the prior distribution $\pi(\bm{\theta}|M)$ to a function $f_{\lambda}(\bm{\theta}$) that is proportional to the posterior distribution. This is done through iteratively constructing a series of intermediate probability distributions:
\begin{equation}
\begin{gathered}
f_j(\bm{\theta}) \sim [p(\bm{D}|\bm{\theta},M)]^{q_j}\cdot\pi(\bm{\theta}|M), j = 0, \hdots,\lambda\\
0 = q_0 < q_1 < \hdots < q_{\lambda} = 1.
\end{gathered}
\label{eq:iterativeDist}
\end{equation}

Algorithm \ref{alg:TMCMC} begins by taking $N_0$ samples $\bm{\theta}_{0,k}$ from the prior distribution $f_0(\bm{\theta}) = \pi(\bm{\theta}|M)$. Then, for each stage $j$, the current samples are used to compute the plausibility weights $w(\bm{\theta}_{j,k})$, i.e. the likelihood of each sample, as
\begin{equation*}
w(\bm{\theta}_{j,k}) = \dfrac{f_{j+1}(\bm{\theta}_{j,k})}{f_j(\bm{\theta}_{j,k})} = [p(\bm{D}|\bm{\theta}_{j,k},M)]^{q_{j+1} - q_j}.
\end{equation*}
The $q_{j+1}$ are selected iteratively in order to have smooth transitions from the intermediate distributions to the posterior distribution. Recent literature suggests that it should make the covariance of the plausibility weights at stage $j$ smaller than a tolerance, often 1.0 \cite{Hadjidoukas15, Ching07}.

The algorithm next computes $S_j$, the average of the plausibility weights, the normalized plausibility weights $\overline{w}(\bm{\theta}_{j,k})$, and the scaled covariance $\overline{\bm{\Sigma}}_j$ of the samples $\bm{\theta}_{j,k}$, which is used to produce the next generation of samples $\bm{\theta}_{j+1,k}$: 
\begin{equation}
\begin{gathered}
S_j = \frac{1}{N_j}\sum_{k=1}^{N_j}w(\bm{\theta}_{j,k})\\
\overline{w}(\bm{\theta}_{j,k}) = w(\bm{\theta}_{j,k})/\sum_{k=1}^{N_j}w(\bm{\theta}_{j,k}) = w(\bm{\theta}_{j,k})/(N_jS_j)\\
\overline{\bm{\Sigma}}_j = b^2\sum_{k=1}^{N_j}\overline{w}(\bm{\theta}_{j,k})[\bm{\theta}_{j,k} - \bm{\mu}_j][\bm{\theta}_{j,k} - \bm{\mu}_j]^T.
\end{gathered}
\label{eq:iterativeDist}
\end{equation}
$\overline{\bm{\Sigma}}_j$ is calculated using the sample mean $\bm{\mu}_j$ and a scaling factor $b$, usually 0.2 \cite{Hadjidoukas15, Ching07}.

$N_{j+1}$ samples $\hat{\bm{\theta}}_{j+1,k}$ are generated by randomly sampling from the previous generation of samples $\{\bm{\theta}_{j,k}\}$ such that $\hat{\bm{\theta}}_{j+1,\ell} = \bm{\theta}_{j,k}$ with probability $\overline{w}(\bm{\theta}_{j,k})$. Note that these parameters are sampled uniformly at random, so any given set can be chosen multiple times -- call $n_{j+1,k}$ the number of times $\bm{\theta}_{j,k}$ is chosen. Each unique sample is used as the starting point of an independent Markov chain of length $n_{j+1,k}$ generated using the Metropolis algorithm with target distribution $f_j$ and a Gaussian proposal distribution with covariance $\overline{\bm{\Sigma}}_j$ centered at the current value. Once the samples have been generated, the algorithm either moves forward to generation $j+1$ or terminates if $q_{j+1} > 1$.

\begin{algorithm}[H]
\caption{TMCMC} \label{alg:TMCMC}
\begin{algorithmic}[1]
\Procedure{TMCMC}{} Ref. \cite{Hadjidoukas15}
\BState BEGIN, SET $j = 0, q_0 = 0$
\BState \textbf{Generate} $\{\bm{\theta}_{0,k}, k = 1, \hdots, N_0\}$ from prior $f_0(\bm{\theta}) = \pi(\bm{\theta}|M)$ and compute likelihood $p(\bm{D}|\bm{\theta}_{0,k}, M)$ for each sample. 

\BState \emph{loop}:
\BState \textbf{WHILE} {$q_{j+1} \leq 1$} \textbf{DO:}
\State \textbf{Analyze} samples $\{\bm{\theta}_{j,k}, k = 1, \hdots, N_j\}$ to determine $q_{j+1}$, weights $\overline{w}(\bm{\theta}_{j,k})$, covariance $\overline{\bm{\Sigma}}_j$, and estimator $S_j$ of $\mathbb{E}[w(\bm{\theta}_{j,k})]$.
\State \textbf{Resample} based on samples available in stage $j$ using the plausibility weights and the Metropolis algorithm in order to generate samples for stage $j+1$ and compute likelihood $p(\bm{D}|\bm{\theta}_{j+1,k},M)$ for each.
\If {$q_{j+1} > 1$}
\State BREAK,
\Else 
\State $j = j+1$
\State \textbf{goto} \emph{loop}.
\EndIf
\State \textbf{end}
\BState \textbf{END}
\EndProcedure
\end{algorithmic}
\end{algorithm}

\section{Results}\label{sec:results}
We now apply the Bayesian framework of the previous section to the cases of a single micro-swimmer and two interacting micro-swimmers. Previous studies have shown complex relationships between fluid resistance and stiffness of the flagellum on emergent swimming speeds and trajectories  \cite{Ho16,Ho19,Leiderman16,Olson15}. 
For both of these examples, we take 2000 samples at each generation of TMCMC. The forward model for the micro-swimmers uses a time-step of $\triangle t$ (refer to Table \ref{Tab:param}) to propagate the solution forward in time up to nondimensional time $T=20$ (corresponding to 40 beat cycles of the tail and 2 seconds of real time). Experimental data are generally in terms of images that are taken in the range of 20 to 60 Hz and resolution often only allows tracking of the center of the head \cite{Mortimer00}. Our \textit{in silico} reference data 
are either a swimming speed based on the center of the head of the swimmer or the distance between the heads and tails of swimmers (in the multi-swimmer case).  To estimate quantities of interest, we utilize observations at $\sim$25 Hz. As a proof of concept, we use simulated observation data by perturbing model outputs with Gaussian noise as
\begin{equation*}
    D_k = \xi_k + \sigma\epsilon_k
\end{equation*}
where $D_k$ is the observation from the $k^{th}$ position of the vector, $\xi_k$ is the $k^{th}$ model output, $\epsilon_k$ is a zero-mean, unit-variance Gaussian variable, and $\sigma$ is the level of the noise. In order to have a meaningful noise-level, we choose $\sigma$ to be a fraction $\sigma = \gamma\beta$, where $\gamma$ is the percent of noise considered, and $\beta$ is the mean of all model outputs. The model prediction error covariance $\bm{\Sigma}$ from Equation \ref{eq:fitness} is assumed to be diagonal, such that $\bm{\Sigma} = \sigma\bm{I}$ whose non-zero entries all have the same magnitude $\sigma$.

\subsection{Single Micro-Swimmer}\label{oneswimmer} 
\begin{figure}[htb]
	\begin{center}
		\begin{tikzpicture}
		\node (z1q) at (-1,0) {\includegraphics[width=2.5in]{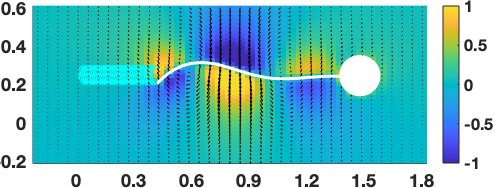}};
		\node (z2q) at (6.3,-1.45) {\includegraphics[width=0.4\textwidth]{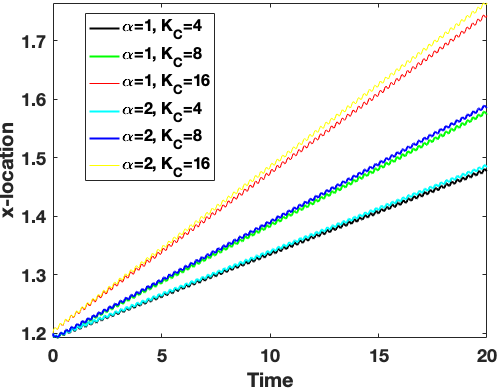}};
		\node (z3q) at (-1,-3) {\includegraphics[width=2.5in]{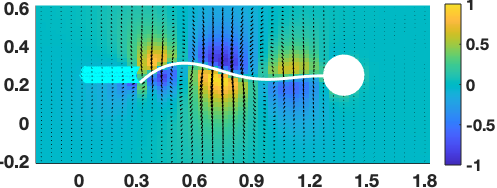}};
		\node (a) at (-1.0,-1.5) {(A) $\alpha=1$,$K_C=8.0$};
		\node (c) at (6.5,-4.4) {(C) Head Center Location };
		\node (b) at (-1.0,-4.6) {(B) $\alpha=1$,$K_C=4.0$  };
		\end{tikzpicture}			
		\end{center}	
\caption{Representative computational model results for the  micro-swimmer. A swimmer at $T=20$ using  resistance parameter $\alpha=1$ and tail curvature stiffness $K_C=8.0$ in (A) and $\alpha=1$ and $K_C=4.0$ in (B). Flow field (arrows) and pressure (colorbar) in (A)-(B) are normalized by the maximum values in (A). The trace of the end of the tail for $T=0-20$ is also shown. A comparison of the $x$-location of the center of the swimmer head is in (C) for different values of $\alpha$ and $K_C$. }\label{fig:SingleRes}
\end{figure}
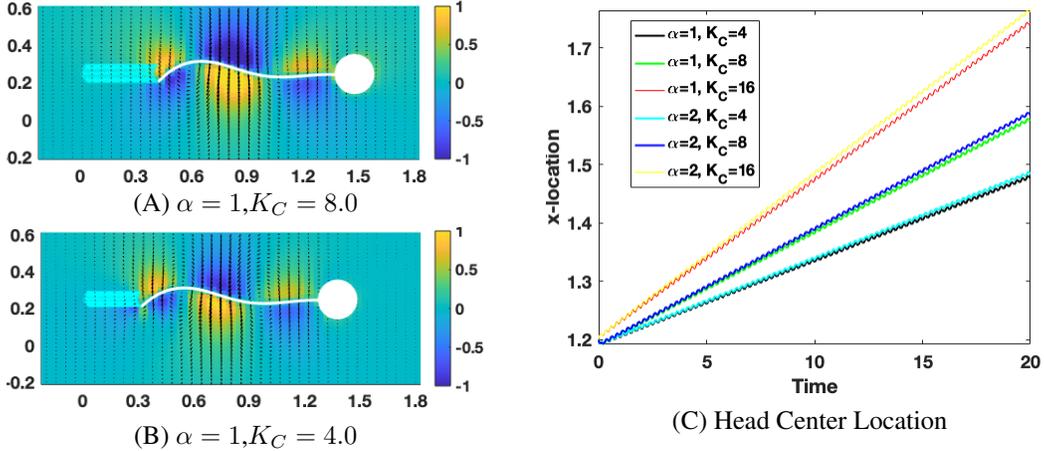

As illustrated in Fig.~\ref{fig:SingleRes}, even though each simulation has the same initial conditions and beat form parameters (Table \ref{Tab:param}), the emergent swimming speed and trajectory of the deterministic swimmer model is a function of both the resistance parameter $\alpha$ (corresponding to the volume fraction of stationary fibers) as well as the parameters relating to the stiffness of the flagellum that is actively bending to propel the swimmer forward. Asymptotic analysis of infinite length swimmers with a prescribed beat form in a Brinkman fluid provides insight into these relationships; increased swimming speeds can be obtained with increased amplitude and/or increased resistance \cite{Leshansky09,Ho16}. However, the work required to achieve larger amplitude bending at a higher volume fraction of fibers in the fluid (corresponding to a larger $\alpha$) becomes too large of a hurdle for micro-swimmers to overcome \cite{Ho16,Ho19}. That is why in models with emergent beat forms, there is often an enhancement in swimming speeds for a moderate range of $\alpha$ and then a decreased swimming speed due to a decreased achieved amplitude at larger $\alpha$ \cite{Leiderman16,Ho19}. In addition, the parameters related to the tail stiffness will lead to a range of emergent beat forms, with different achieved amplitudes, which then results in  different swimming speeds.
As can be seen from Fig.~\ref{fig:SingleRes}(A) to (B), halving only the tail curvature stiffness parameter $K_C$ results in a slower swimmer with smaller achieved amplitude (seen in tail traces). From Eq.~\eqref{EFB}, we can observe that the magnitude of $K_C$ controls how closely the swimmer will try to maintain the preferred beat form. In (C), the $x$-location of the center of the head can be seen for different $\alpha$ and $K_C$. At this range of parameter values, we observe that the higher $\alpha$ results in a small increase in forward progression and that the larger $K_C$ results in a larger increase in forward progression. The small oscillations in the graphs correspond to the beat form where the swimmer may actually move backwards for a small portion of the beat but in the overall beat, forward progression is made.

For the single micro-swimmer case, the model outputs $\bm{g}$ are generated from our deterministic model. Using the center of the head of the micro-swimmer, the $x$- and $y$- velocities are estimated, resulting in 50 $x$-velocity and 50 $y$-velocity observations when solving up to $T=20$ (data at 25 Hz). To illustrate the feasibility of this approach, we consider two scenarios for parameter estimation: first, estimating only the fluid resistance $\alpha$ and the noise level $\sigma/\beta$  and second, estimating the fluid resistance $\alpha$, the tail curvature stiffness parameter $K_C$, and the noise level $\sigma/\beta$. The nominal parameter values taken are $\alpha = 1.0$ and $K_C = 8.0$. In both of these cases, we also demonstrate the strength of the method by perturbing the observation data by 1\%, 10\%, and 20\% noise.

\begin{figure}
\begin{center}
		\begin{tikzpicture}
		\node (z1q) at (-6.9,0) {\includegraphics[width=2.8in]{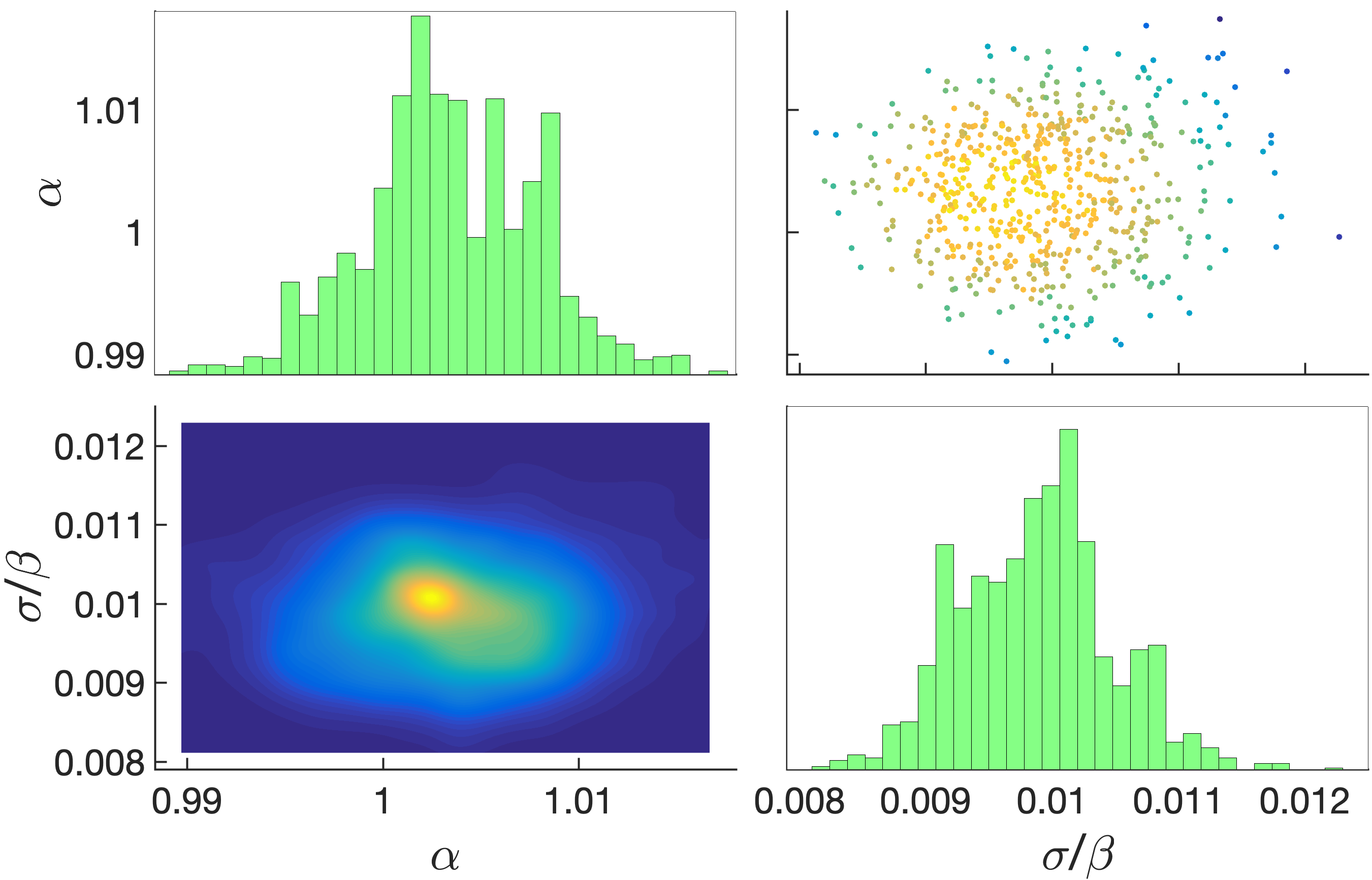}};
		\node (z2q) at (1.4,0) {\includegraphics[width=3.0in]{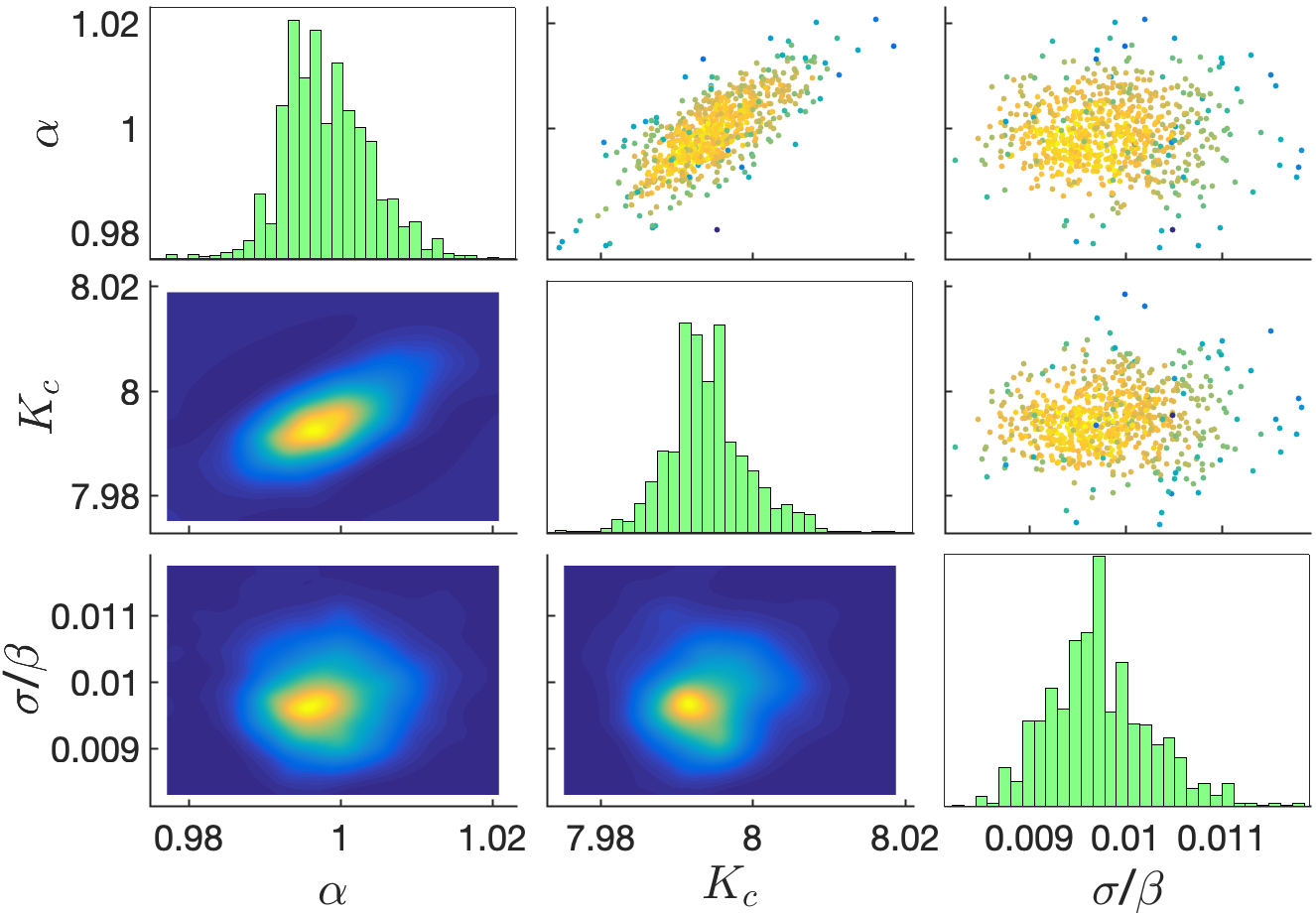}};
		\node (z3q) at (-6.9,-6.2) {\includegraphics[width=2.8in]{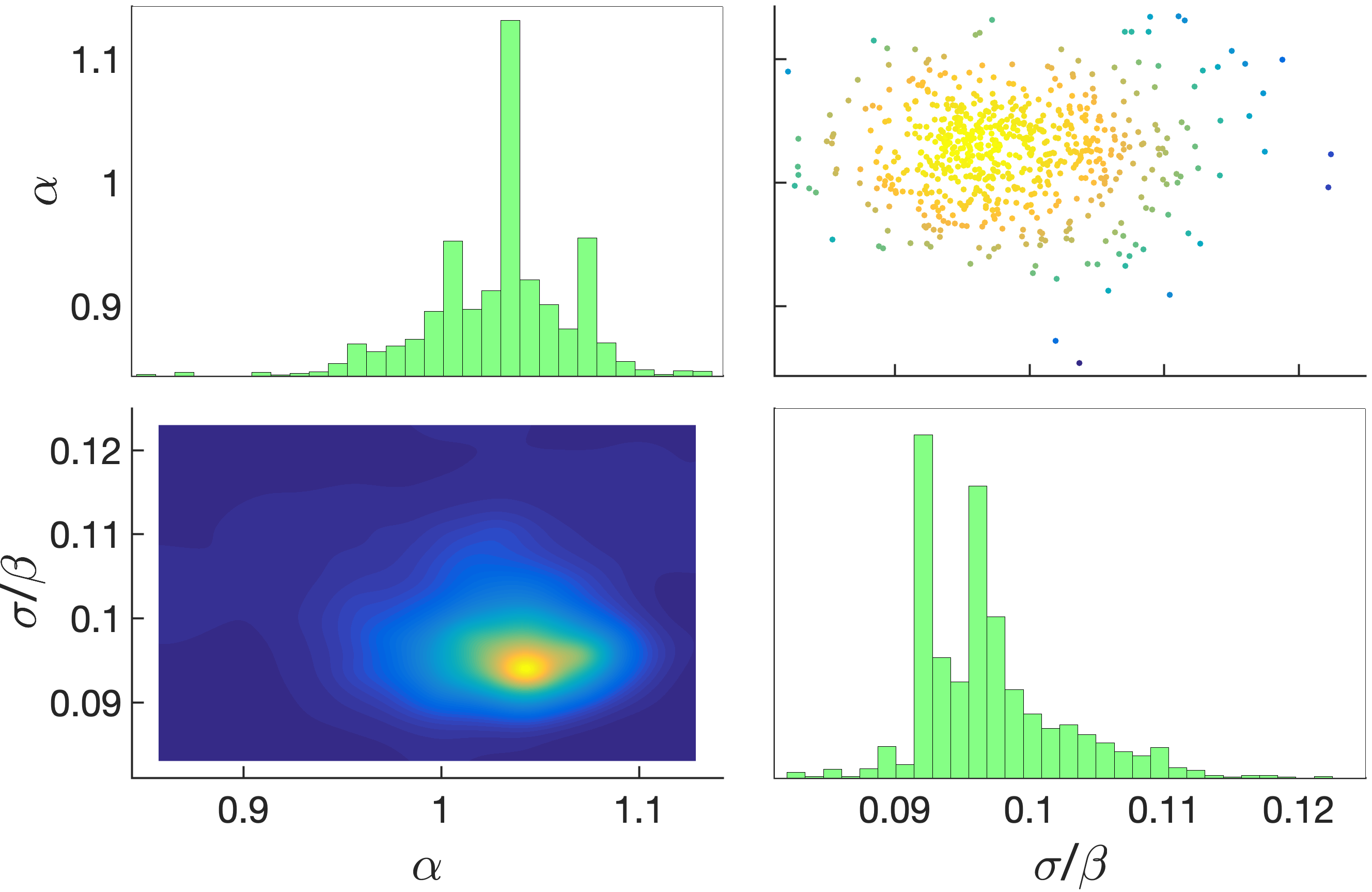}};
		\node (z4q) at (1.4,-6.2) {\includegraphics[width=3.0in]{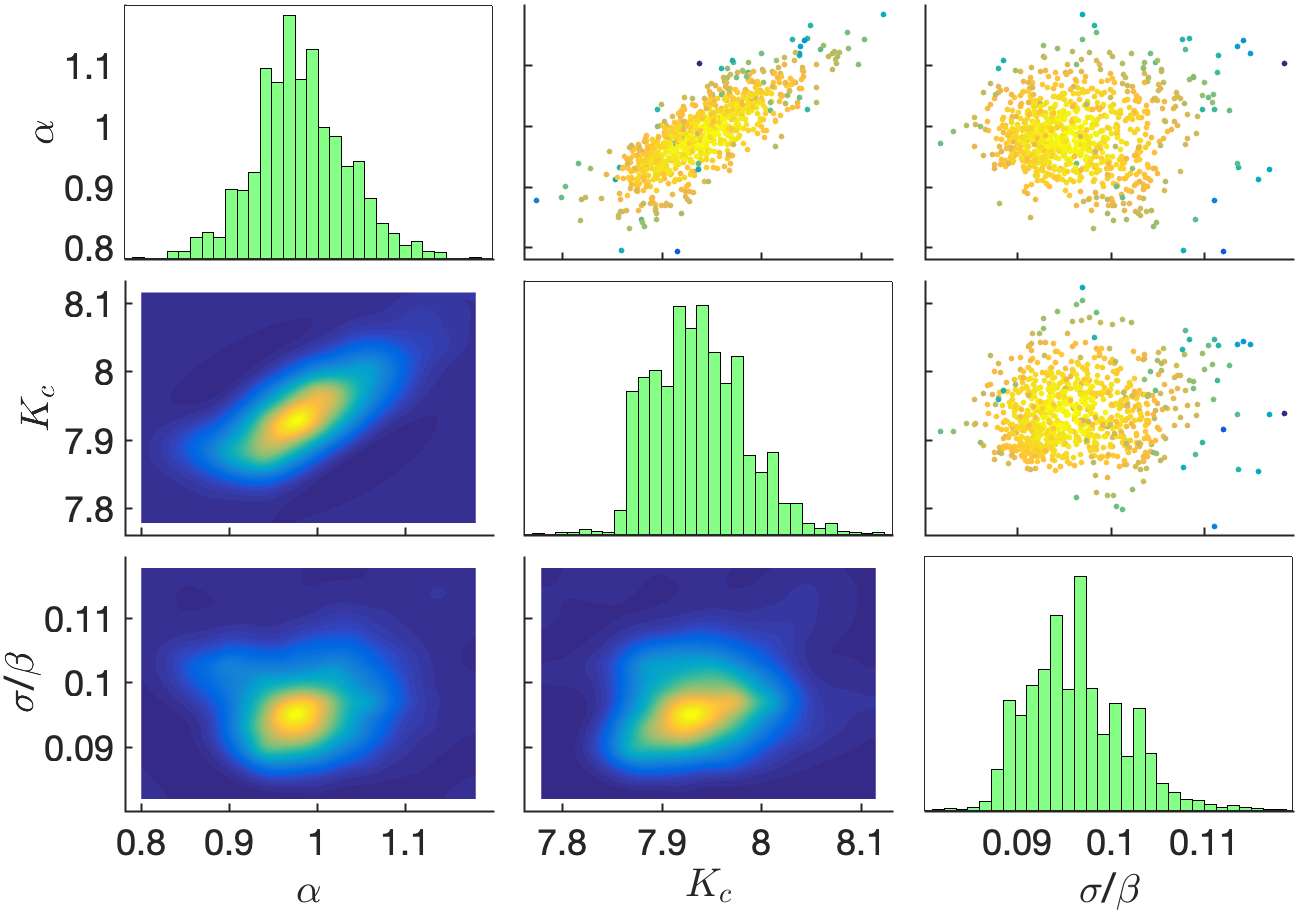}};
		\node (z3q) at (-6.9,-12.3) {\includegraphics[width=3.0in]{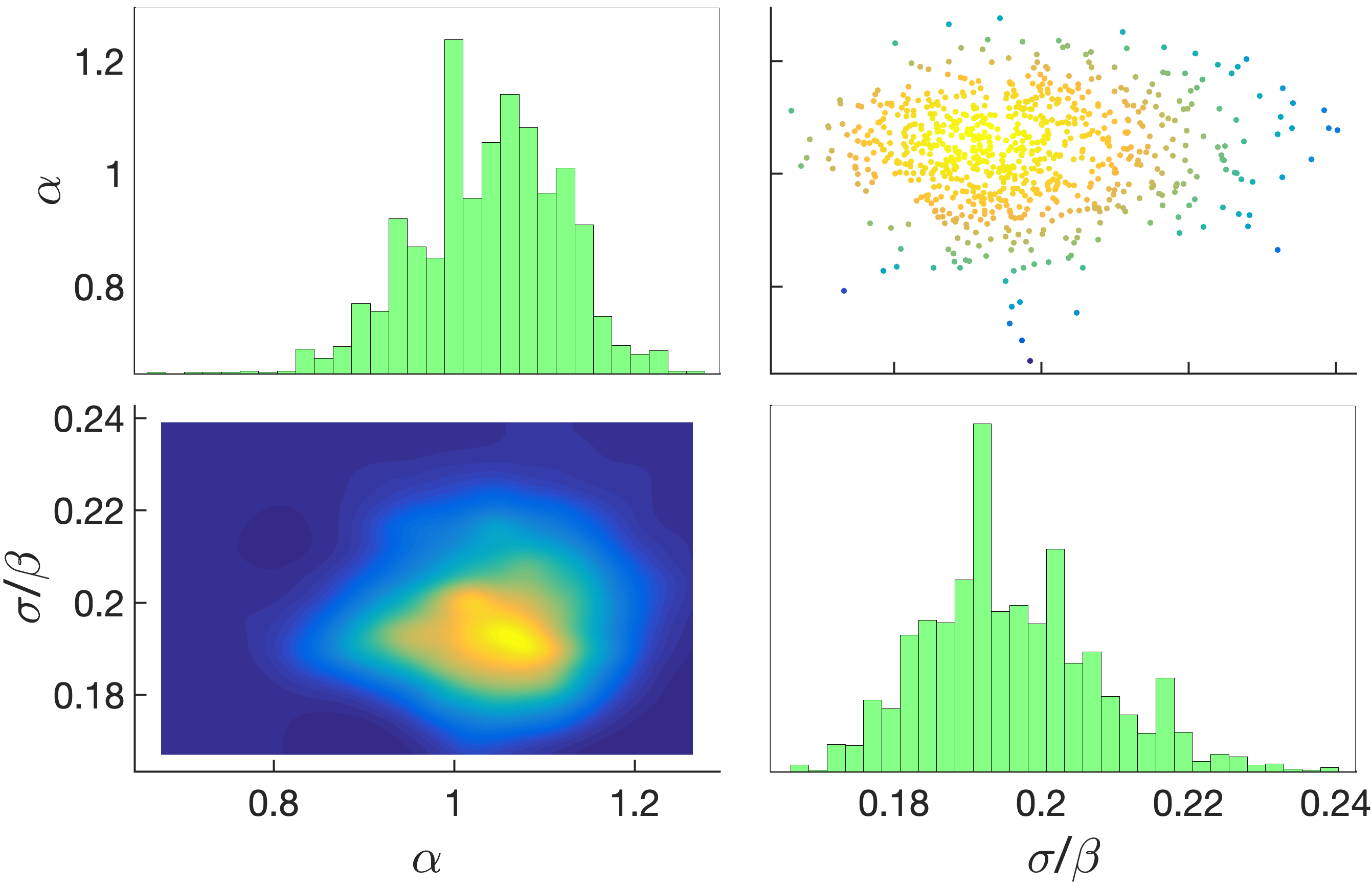}};
		\node (z4q) at (1.4,-12.3) {\includegraphics[width=3.1in]{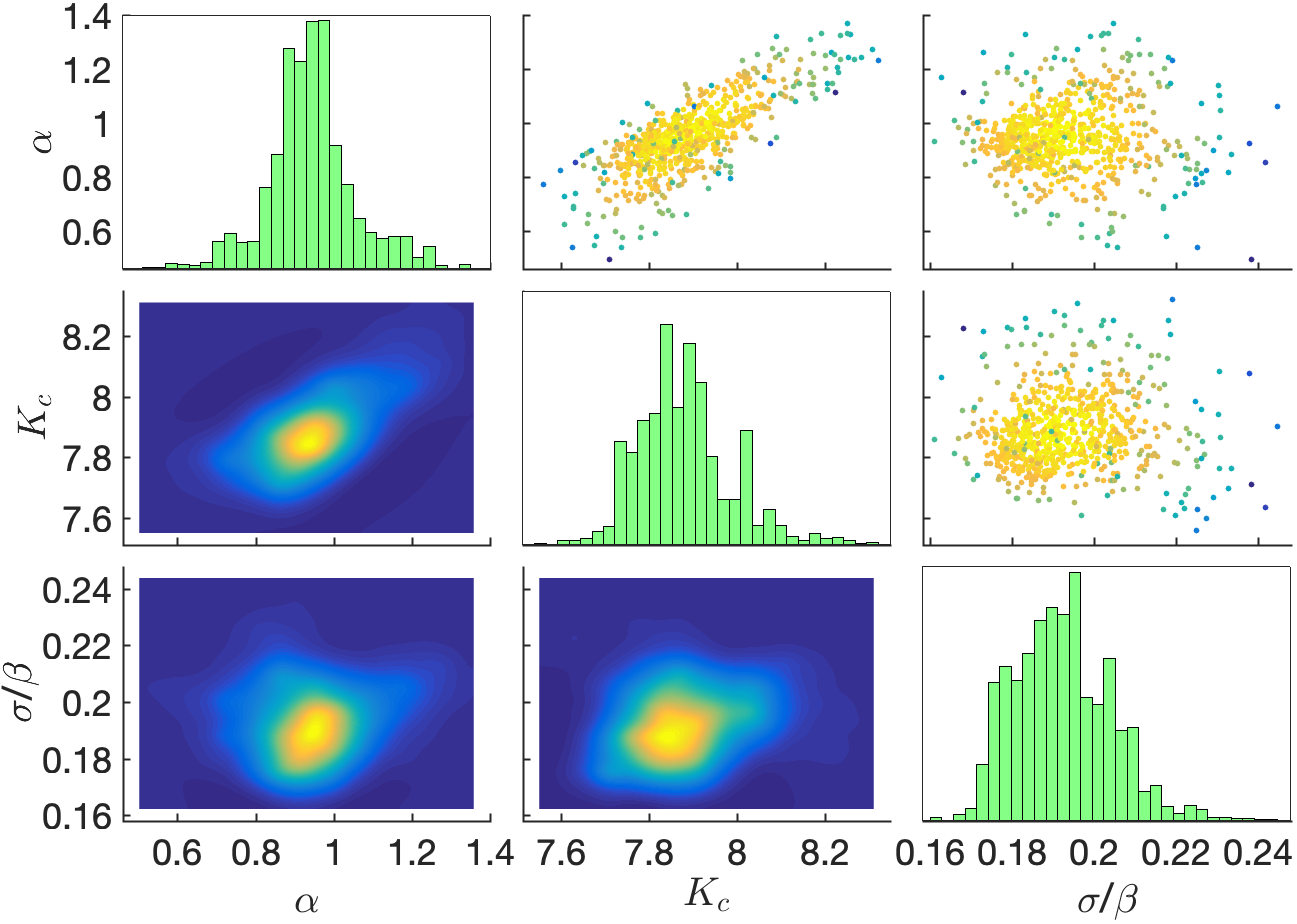}};
		\node (a) at (-6.8,-3.0) {(A) One Parameter, 1\% noise};
		\node (d) at (1.8,-3.0) {(D) Two Parameters, 1\% noise};
		\node (b) at (-6.8,-9.2) {(B) One Parameter, 10\% noise};
		\node (e) at (1.8,-9.2) {(E) Two Parameters, 10\% noise};
		\node (c) at (-6.8,-15.4) {(C) One Parameter, 20\% noise};
		\node (f) at (1.8,-15.4) {(F) Two Parameters, 20\% noise};
		\end{tikzpicture}			
		\end{center}	
\caption{Parameter estimation results for a single swimmer with fluid resistance $\alpha = 1.0$ and tail curvature stiffness $K_C = 8.0$. In each experiment, noisy velocity data from the center of the head were used to track the movement. For each noise level, the experiment was run until $T = 20$. Histograms for each parameter are displayed along the main diagonal of the figure. Subfigures below the diagonal show the marginal joint density functions for each pair of parameters, while subfigures above the diagonal show the samples used in the final stage of TMCMC. Colors correspond to probabilities, with yellow likely and blue unlikely.}
\label{fig:one_swimmer}
\end{figure}
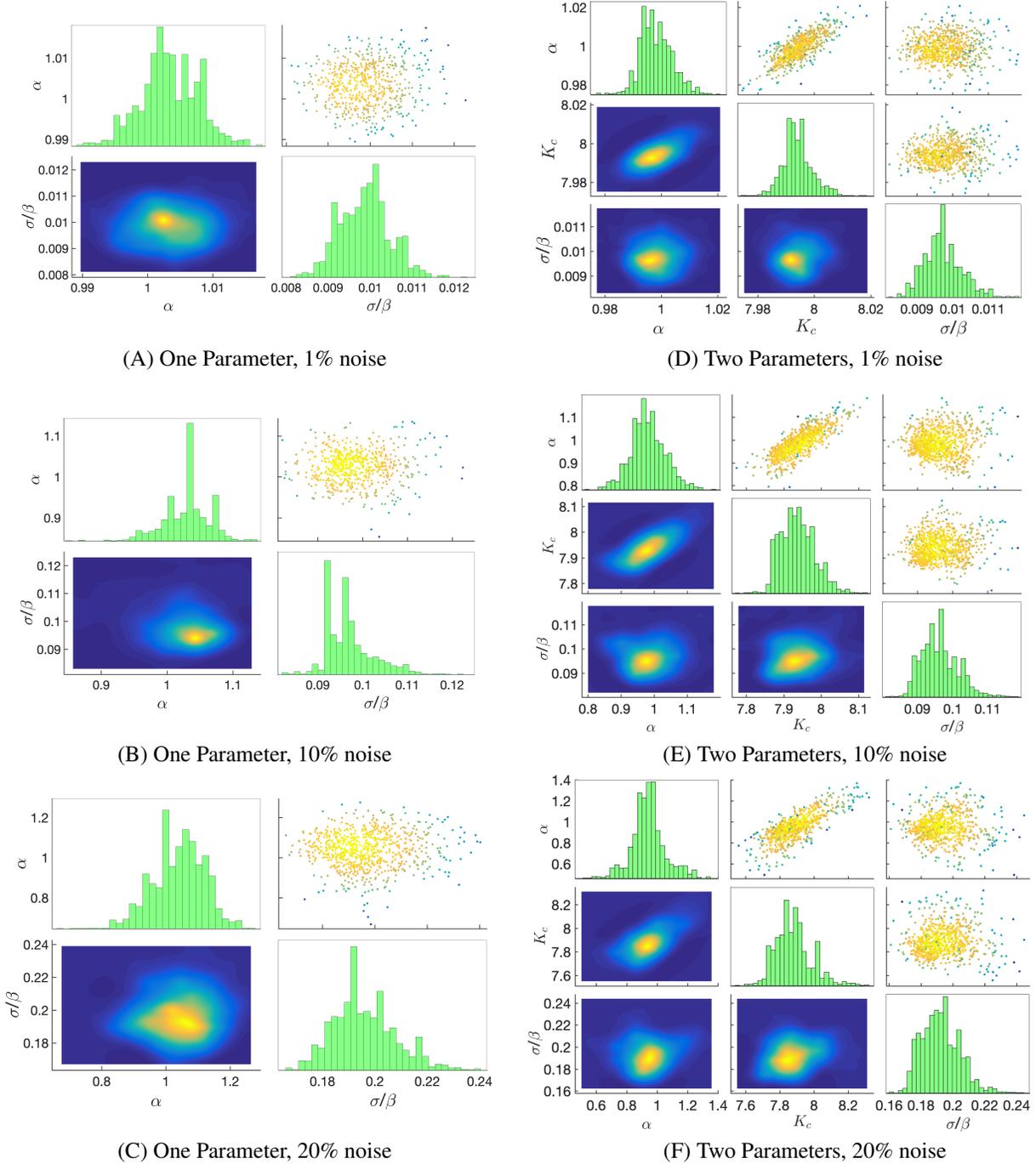

\begin{table}[h]
\centering
\begin{tabular}{c c c c c } 
 \hline
 \hline
 Noise Level & $\alpha$ & $u_{\alpha}$ (\%) & $\sigma/\beta$ & $u_{\sigma/\beta}$ (\%)  \\
 \hline
1\% & 1.0035 &0.436 &0.00987 &6.006  \\
10\% &1.0297 &3.354 &0.0968 &5.376 \\
20\% &1.0421 & 7.780&0.196 &6.006  \\
 \hline
\end{tabular}\smallskip
\caption{Posterior means and uncertainties for parameter estimation on $\alpha$ for the single swimmer using velocity data for varying noise levels.}
\label{table:one_swim_onep}
\end{table}

For the first case, we assume a uniform prior for the parameters $(\alpha, \sigma/\beta)$ on the space $[0.0, 2.0] \times [0.0, 0.5]$.  The results are summarized in Table \ref{table:one_swim_onep}, and the 1\% noise case is displayed in Fig.~\ref{fig:one_swimmer}(A). In this case, the joint marginal distribution is an ellipsoid, matching intuition as the noise and the resistance of the fluid $\alpha$ should be uncorrelated. Considering the recovered parameter values in row one of Table \ref{table:one_swim_onep}, the parameters found are 1.0035 for $\alpha$ and 0.00987 for $\sigma/\beta$, matching closely to the nominal values of 1 and 0.01, respectively. To quantify the degree of uncertainty for each parameter's posterior distribution, we compute the coefficient of variation, defined as the ratio between a parameter's standard deviation to its mean (denoted by $u_{\alpha}$ and $u_{\sigma/\beta}$ here). In this case, the coefficient of variation for $\alpha$ is 0.436\%, showing that we are fairly certain in the recovered mean. In addition, the nominal parameter values are included within two-standard deviations of the recovered means for both parameters.  

The same experiment is run using 10\% and 20\% additive Gaussian noise, displayed in Figs.~\ref{fig:one_swimmer}(B) and (C) and Table \ref{table:one_swim_onep}. For both of these cases, we see that the distributions include the nominal parameter values. However, the distributions are much more spread out than the 1\% noise case. For example, the 1\% additive noise joint posterior distribution for $\alpha$ ranges in value from 0.99 to 1.01, while the 10\% additive noise ranges from 0.9 to 1.1 and the 20\% noise ranges from 0.8 to 1.2. This relationship is also seen in the coefficient of variation for the $\alpha$ parameter: $u_\alpha$ for 10\% additive noise is about 10 times as large as that for 1\% noise and the 20\% noise is approximately double that of the 10\% noise, matching very closely to the increase in the noise level. In addition, we can see that the coefficient of variation for the noise level stays the same between the various levels, matching intuition as the certainty in the noise should not change as more noise is added to the system.

 \begin{table}[h]
\centering
\begin{tabular}{c c c c c c c} 
 \hline
 \hline
 Noise Level & $\alpha$ & $u_{\alpha}$ (\%) &$K_C$ & $u_{K_C}$ (\%)& $\sigma/\beta$ & $u_{\sigma/\beta}$ (\%)  \\
 \hline
1\% & 0.9982 &  0.580 &7.9944 & 0.0647& 0.009728 &5.47  \\
10\% &0.9820 & 5.47 &7.9368 &0.58 &0.0964 &5.34 \\
20\%& 0.9426 & 11.93 & 7.8800 & 1.35 & 0.1922 & 6.11\\

 \hline
\end{tabular}\smallskip
\caption{Posterior means and uncertainties for parameter estimation on $\alpha$ and $K_C$ for the single swimmer using velocity data for varying noise levels.}
\label{table:one_swim_twop}
\end{table}

Next, we estimate both $\alpha$ and tail curvature stiffness parameter $K_C$, assuming a uniform prior on $[0, 2] \times [0, 16] \times [0, 0.5]$ Again, we also perturb the data using 1\%, 10\%, and 20\% Gaussian noise.
In the results displayed in Table \ref{table:one_swim_twop} and Fig.~\ref{fig:one_swimmer}(D) for the 1\% noise level, we can see that there is a positive correlation between the fluid resistance parameter $\alpha$ and tail curvature stiffness $K_C$: as the  volume fraction of fibers in the fluid increases (larger $\alpha$), a stiffer tail is required to achieve  similar swimming speeds. Like the one-parameter case, the joint marginal distributions including the noise have a single mode and are uncorrelated, i.e. the parameters are independent of the noise.

We repeated these experiments using 10\% and 20\% additive noise, displayed in Figs.~\ref{fig:one_swimmer}(E) and (F) and Table \ref{table:one_swim_twop}. From the table, we can see that all nominal parameter values are recovered within two-standard deviations of the estimated means. Similar to the single parameter case, the noise for $\alpha$ and $K_C$ also scales with the overall level of additive noise: the 10\% noise case has about 10 times more uncertainty than the 1\% case, and the 20\% noise case has standard deviations twice as large as the 10\% noise case. In all of the figures, there is also a strong positive correlation between $\alpha$ and $K_C$, as seen with 1\% additive noise.

 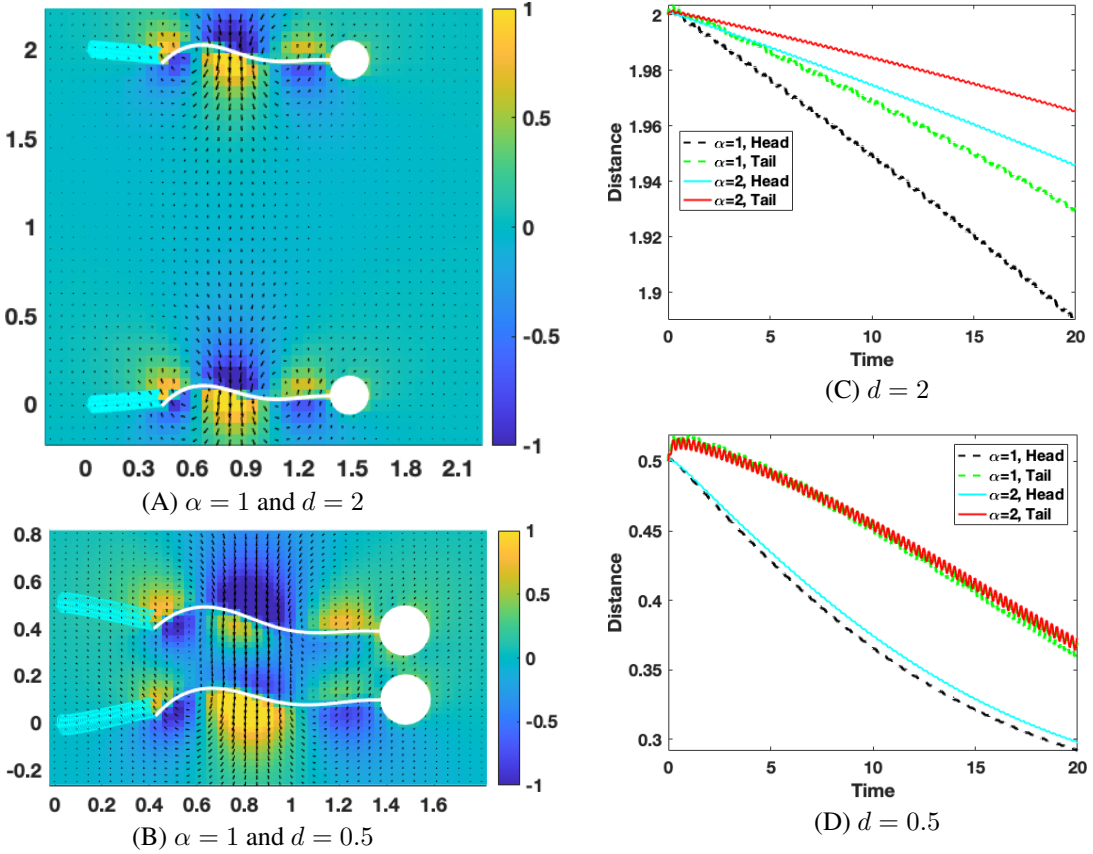
\begin{figure}[htb!]
\begin{center}
		\begin{tikzpicture}
		\node (z1q) at (-1.9,0) {\includegraphics[width=2.9in]{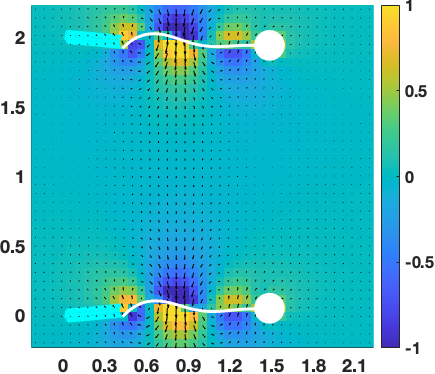}};
		\node (z2q) at (5.6,0.75) {\includegraphics[width=2.5in]{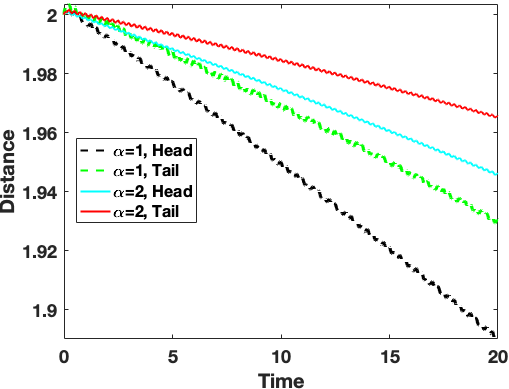}};
		\node (z3q) at (-1.9,-5.7) {\includegraphics[width=2.9in]{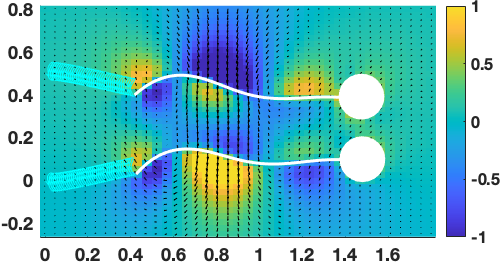}};
		\node (z4q) at (5.6,-4.95) {\includegraphics[width=2.5in]{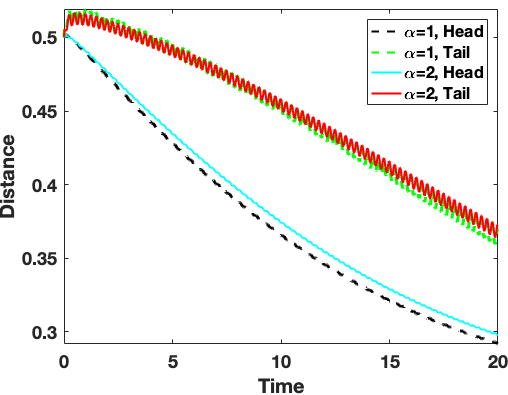}};
		\node (a) at (-2.3,-3.5) {(A) $\alpha=1$ and $d=2$};
		\node (c) at (6.0,-2) {(C) $d=2$};
		\node (b) at (-2.3,-7.95) {(B) $\alpha=1$ and $d=0.5$};
		\node (d) at (6.0,-7.75) {(D) $d=0.5$};
		\end{tikzpicture}			
		\end{center}
\caption{Representative simulation results for two micro-swimmers that are initially a distance of 2 units apart in (A) and 0.25 units apart in (B). These are both for the resistance parameter $\alpha=1$ at $T=20$; the time course of the tail undulations of each swimmer are highlighted to the left of the swimmer and the normalized flow field is depicted with arrows and the pressure is visualized with the colorbar. For each of the cases, the distance between the center of the heads of the swimmers as well as the distance between the last point of each of the swimmers is given in (C) for 2 units apart initially and 0.25 units apart in (D). These are all for tail curvature stiffness $K_C=8.0$.}\label{fig:TwoRes}
\end{figure}

\subsection{Two Micro-Swimmers}\label{twoswimmers} 
Representative results of the deterministic model for the two micro-swimmer case are shown in Fig.~\ref{fig:TwoRes}. Two swimmers are initialized in the same exact initial condition, separated by a vertical distance of $d=2$ in (A) and $d=0.5$ in (B). The flow field and swimmer locations are at $T=20$ and the entire tail trajectory from $T=0-20$ is shown for each swimmer. Each swimmer feels the other swimmer via the fluid and for $\alpha=1$, attraction is observed in both cases, although it is much stronger when initialized at $d=0.5$ in (B). As previously observed, the time scale of attraction will be a function of $\alpha$ as well as the stiffness parameters of the tail \cite{Olson15,Leiderman16}. In (C)-(D), the distance between the head centers and the last points on the tail are shown for initial separations of $d=2$ and $d=0.5$, respectively. 

To investigate parameter estimation and uncertainty quantification in the two micro-swimmer case, the observations at 25 Hz will correspond to two data points: the distance between the center of the two swimmers' heads and the distance between the last points on the tail of the swimmers. The swimmers initially start parallel to each other with the center of their heads $d=2$ units apart and we solve up to $T=20$. We use the same amount of noise on both sets of observations.

\begin{figure}
\begin{center}
		\begin{tikzpicture}
		\node (z1q) at (-6.9,0) {\includegraphics[width=3.1in]{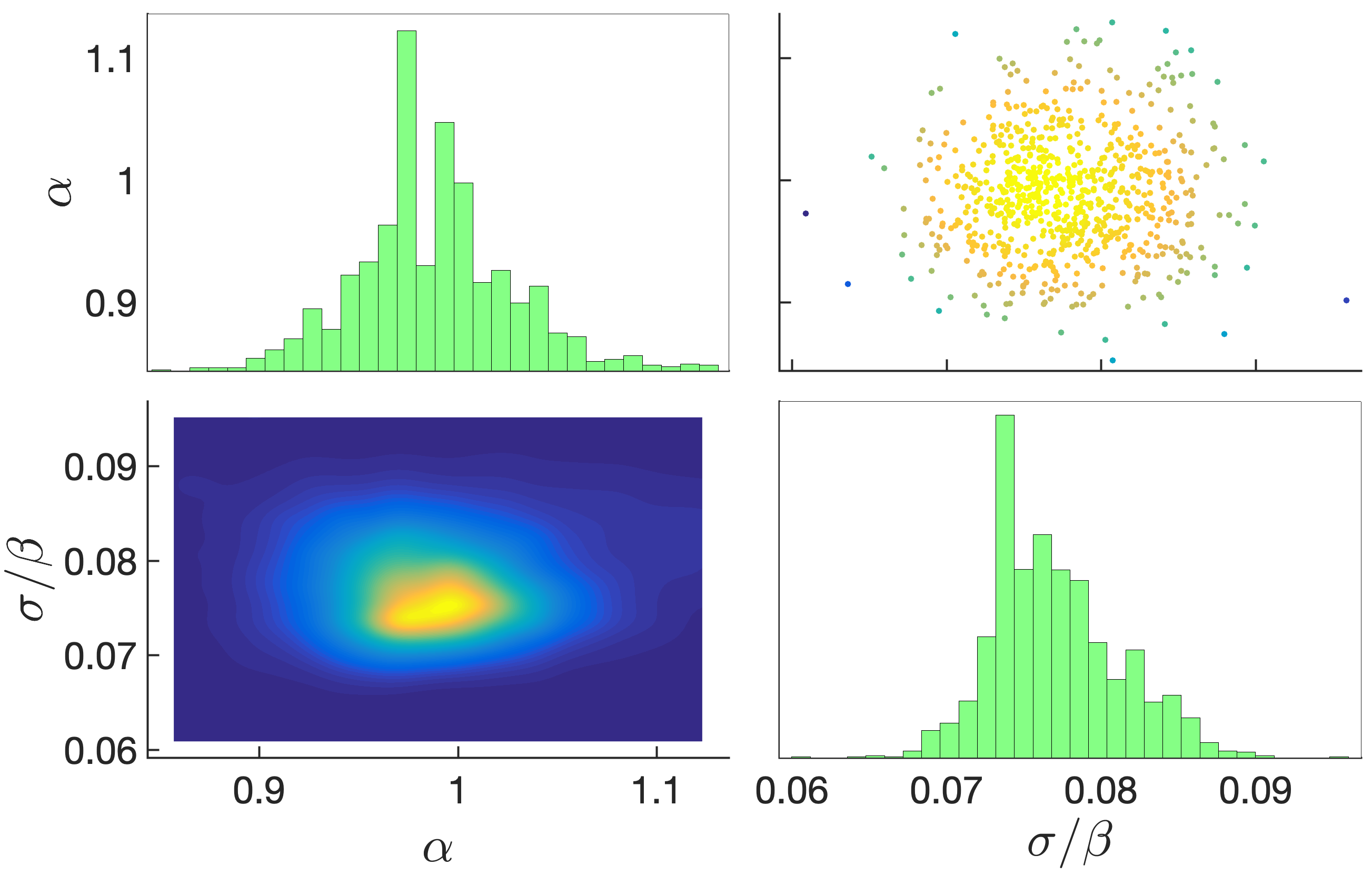}};
		\node (z2q) at (1.4,0) {\includegraphics[width=3.2in]{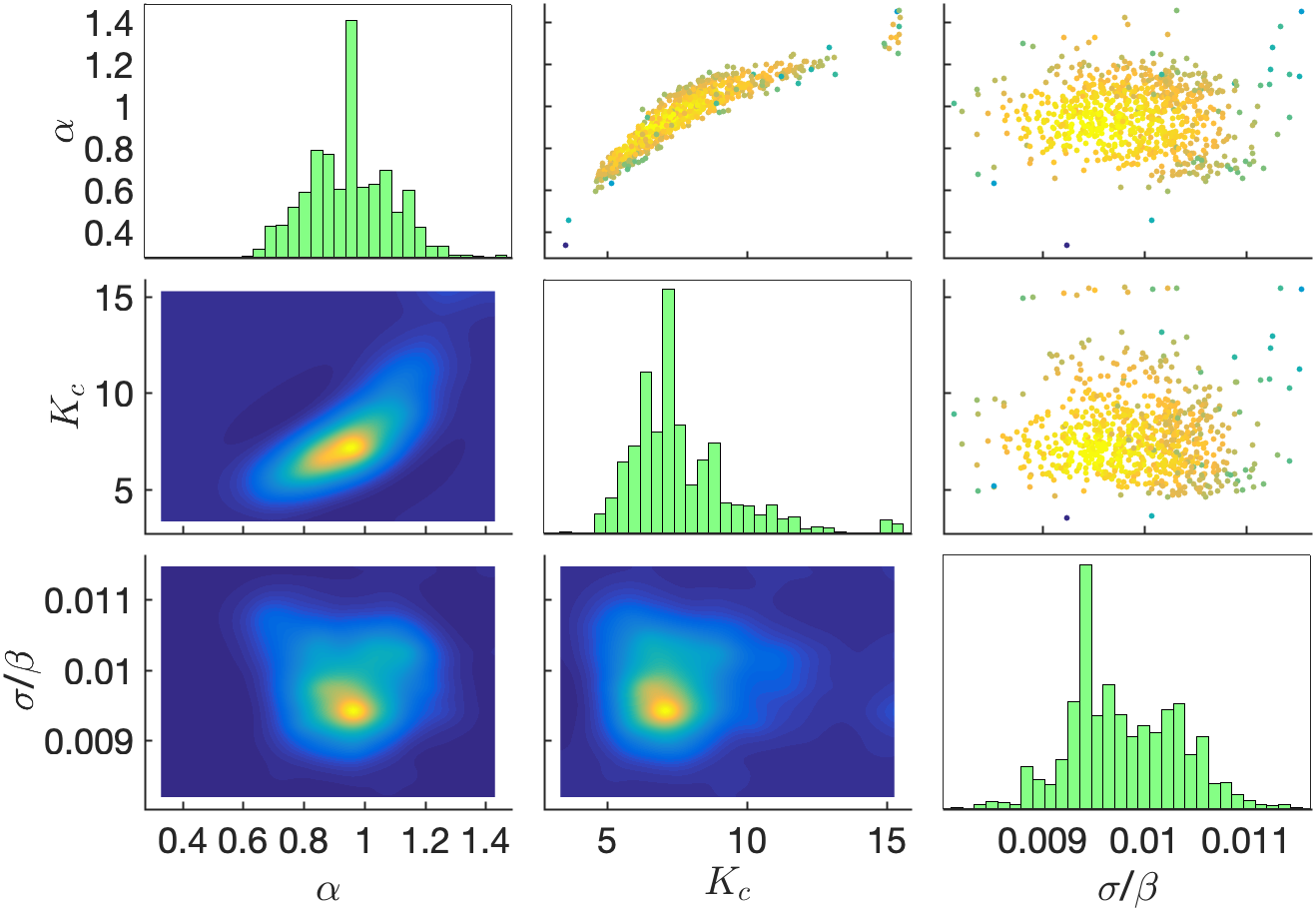}};
		\node (z3q) at (-6.9,-6.5) {\includegraphics[width=3.1in]{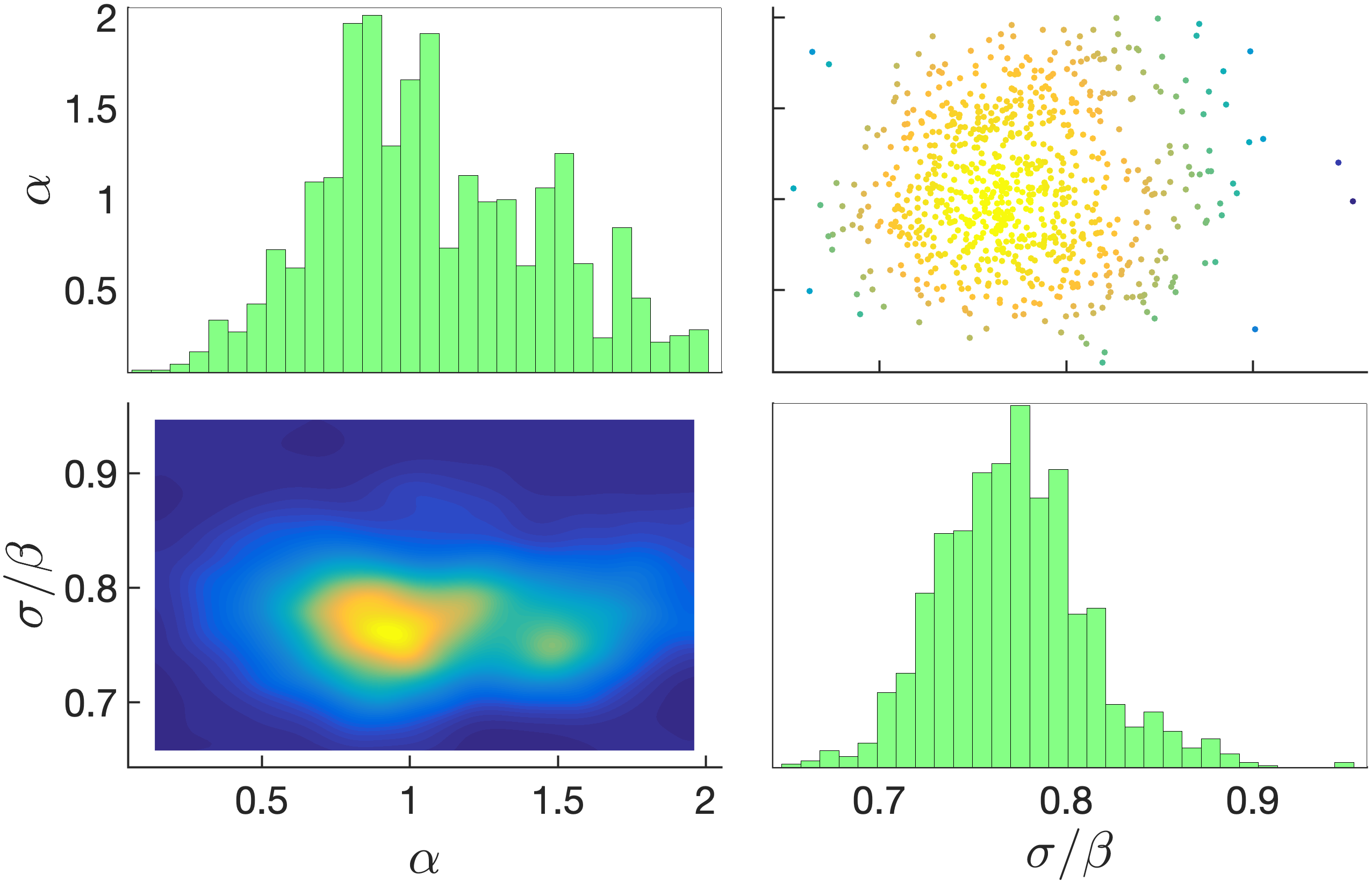}};
		\node (z4q) at (1.4,-6.5) {\includegraphics[width=3.2in]{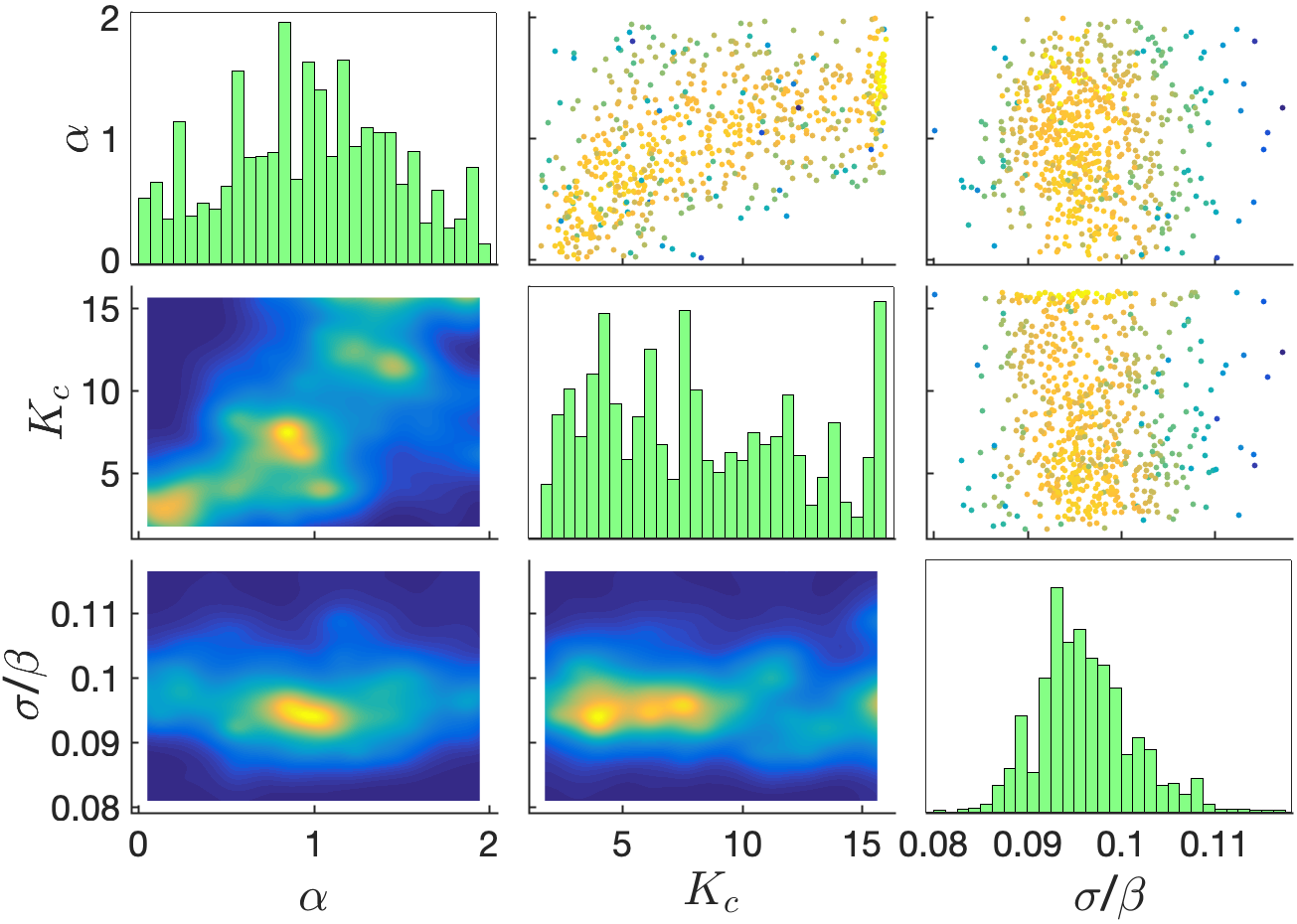}};
		\node (a) at (-6.8,-3.2) {(A) One Parameter, 1\% noise};
		\node (c) at (1.8,-3.2) {(C) Two Parameters, 1\% noise};
		\node (b) at (-6.8,-9.8) {(B) One Parameter, 10\% noise};
		\node (d) at (1.8,-9.8) {(D) Two Parameters, 10\% noise};
		\end{tikzpicture}			
		\end{center}	
\caption{Parameter estimation results for two swimmers with fluid resistance $\alpha = 1.0$ and tail curvature stiffness $K_C = 8.0$. In each experiment, noisy distance data between the center of the heads and tips of the tails were used as reference data. For each noise level, the experiment was run until $T = 20$. Histograms for each parameter are displayed along the main diagonal of the figure. Subfigures below the diagonal show the marginal joint density functions for each pair of parameters, while subfigures above the diagonal show the samples used in the final stage of TMCMC. Colors correspond to probabilities, with yellow likely and blue unlikely.}
\label{fig:two_swimmer}
\end{figure}
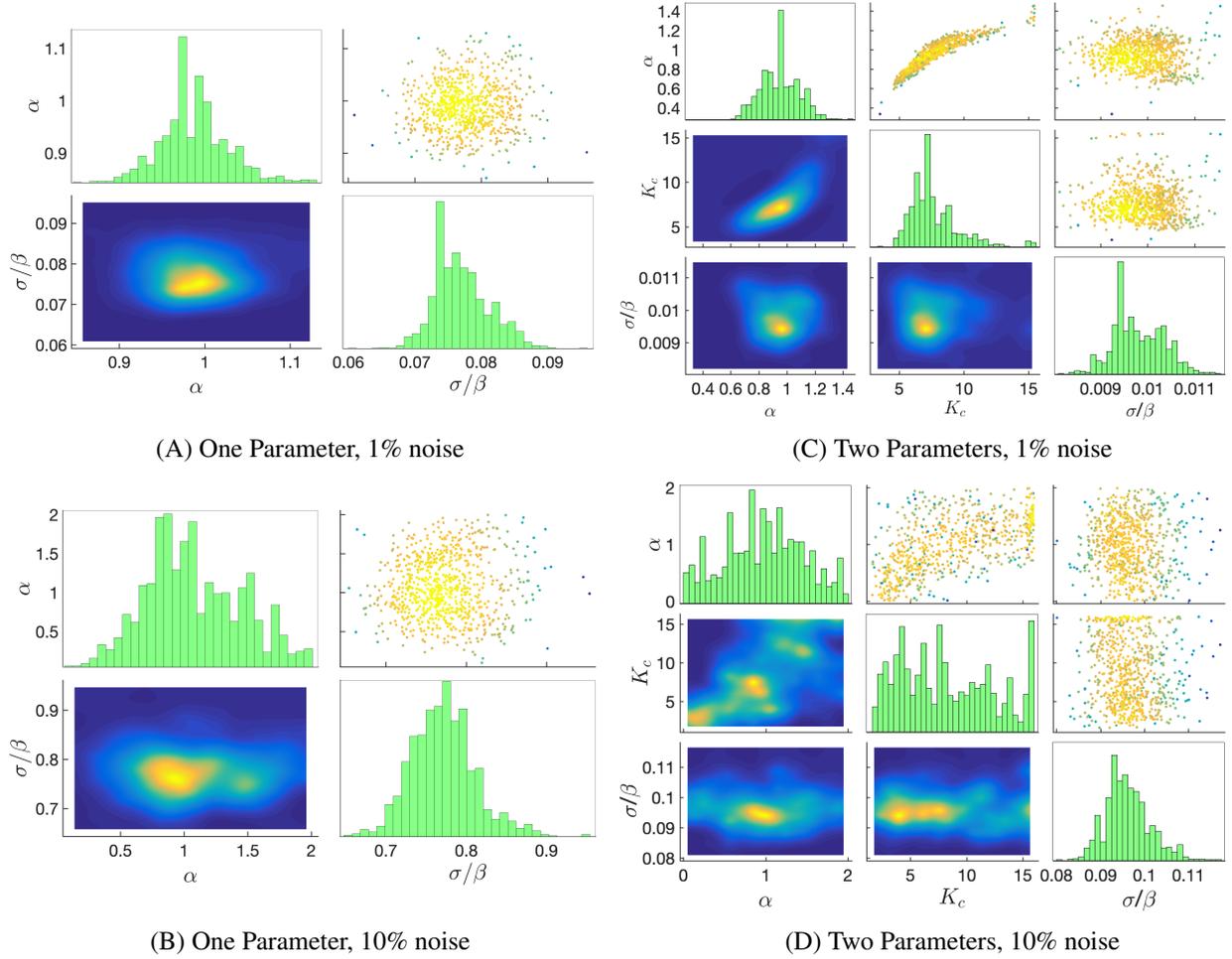

The 1\% noise case using 2000 samples for each generation of TMCMC for estimating $\alpha$ and $\sigma/\beta$ is displayed in Fig.~\ref{fig:two_swimmer}(A). The joint marginal distribution for $\alpha$ and $\sigma/\beta$ is unimodal and uncorrelated. In addition, it is centered closely around the nominal parameter value for $\alpha$ and includes the nominal parameter value for the level of the noise. The distributions for the 10\% noise case are displayed in Fig.~\ref{fig:two_swimmer}(B) and the results are summarized in Table \ref{table:two_swim_onep}. Similar to the 1\% noise case, the joint marginal distribution for $\alpha$ and $\sigma/\beta$ is ellipsoidal, displaying that the two parameters are uncorrelated. The recovered means include the nominal parameter values within one-standard deviation. Furthermore, the coefficient of variation for the parameters increases in proportion to the increase in the noise level: for 1\% noise, the coefficient of variation $u_{\alpha} = 3.94 \%$, while for the 10\% noise case, the coefficient of variation $u_{\alpha} = 34.67\%$. This shows as the noise in the parameters increase, the certainty in our parameter values correspondingly decreases.

 \begin{table}[h]
\centering
\begin{tabular}{c c c c c } 
 \hline
 \hline
 Noise Level & $\alpha$ & $u_{\alpha}$ (\%) & $\sigma/\beta$ & $u_{\sigma/\beta}$ (\%)  \\
 \hline
1\% & 0.9878& 3.94&0.00962&5.35  \\
10\% & 1.0763& 34.67&0.09646 &5.14\\
 \hline
\end{tabular}\smallskip
\caption{Posterior means and uncertainties for parameter estimation on $\alpha$ for two swimmers using distance between the heads and tails as noisy observation data.}
\label{table:two_swim_onep}
\end{table}

As with the single swimmer case, we also estimated both the fluid resistance $\alpha$ and the tail curvature stiffness $K_C$ of the micro-swimmers through observing the distance between the head and tails of two swimmers. The results are summarized in Fig.~\ref{fig:two_swimmer}(C)-(D) and Table \ref{table:two_swim_twop}. As in the single swimmer case, there is a positive correlation between the fluid resistance and the curvature stiffness for the tails of the micro-swimmers. However, we note that the uncertainty in the parameters are much larger than the single-swimmer case, and grows as the noise in the observations increase. One reason for this is that the dynamics of attraction may not yet be achieved by time $T = 20$ because of the distance apart that the swimmers began (refer to Fig.~\ref{fig:TwoRes}(A) and (C)). Due to this and the large amount of noise in the system, the resulting distributions are very spread out in the parameter space. In order to better approach the parameter estimation question in this case, data need to be recorded for a longer amount of time so that the swimmers attract and attain their beat formation.

 \begin{table}[h]
\centering
\begin{tabular}{c c c c c c c } 
 \hline
 \hline
 Noise Level & $\alpha$ & $u_{\alpha}$ (\%) & $K_C$ & $u_{K_C}$ (\%) & $\sigma/\beta$ & $u_{\sigma/\beta}$ (\%)  \\
 \hline
1\% & 0.9533 &14.35   &7.6632 &24.56 & 0.00980&5.46  \\
10\%& 0.9768 &49.46 & 8.2374 &50.27 &0.09616 & 5.27\\

 \hline
\end{tabular}\smallskip
\caption{Posterior means and uncertainties for parameter estimation on $\alpha$ and $K_C$ for two swimmers using distance between the heads and last point on tails as noisy observation data.}
\label{table:two_swim_twop}
\end{table}

\subsection{Two Micro-Swimmers, Two Standard Deviations}\label{twosigma}
The swimmers' tails and the swimmers' head distances can be on different scales (shown in Fig.~\ref{fig:TwoRes}(C) and (D)), so using different standard deviations $\sigma_t$ and $\sigma_h$ for the tail data $x_t$ and head data $x_h$ can more accurately represent how noise is added to the system. Here, the covariance matrix is still assumed to be diagonal, but now $\Sigma_{j,j} = \sigma_t$ for $j = 1, 3, \hdots, m - 1$ and $\Sigma_{j,j} = \sigma_h$ for $j = 2, 4, \hdots, m$. The data are corrupted according to
\begin{equation*}
D_{k} = 
\begin{cases}\xi_k + \sigma_t\epsilon_k,\quad k = 1, 3, \ldots, m-1, \\
		\xi_k + \sigma_h\epsilon_k, \quad k = 2, 4, \ldots, m.
\end{cases}
\end{equation*}
Since the previous case had issues resolving the position of the swimmers in part because the swimmers began too far apart, we initially set the center of the heads to be $d=0.5$ units apart and measure the distance between the heads and tails at 25 Hz. We take 2000 samples at each generation of TMCMC and again compare the results for two different noise levels.

\begin{figure}
\begin{center}
		\begin{tikzpicture}
		\node (z1q) at (-6.9,0) {\includegraphics[width=3.2in]{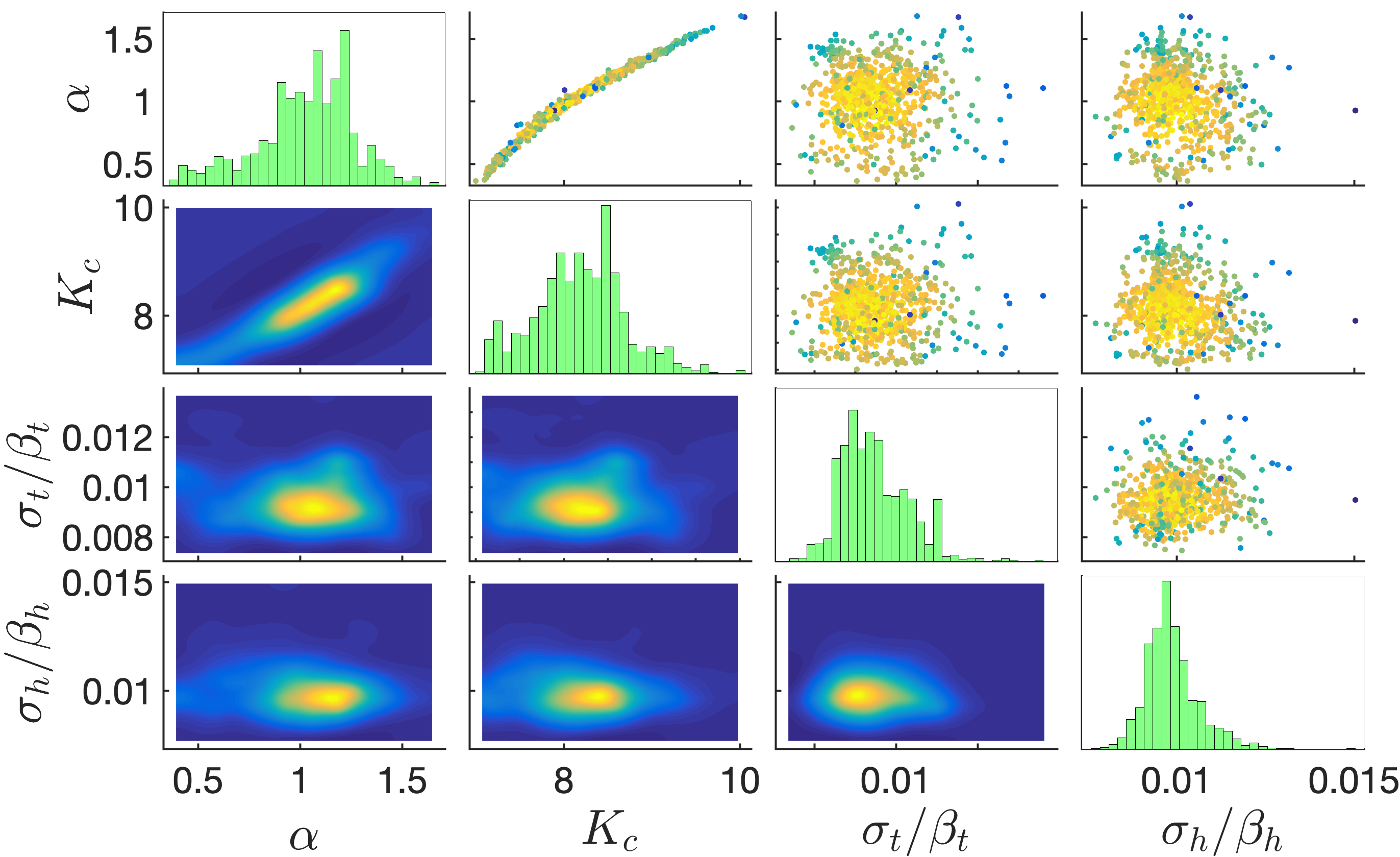}};
		\node (z2q) at (1.4,0) {\includegraphics[width=3.2in]{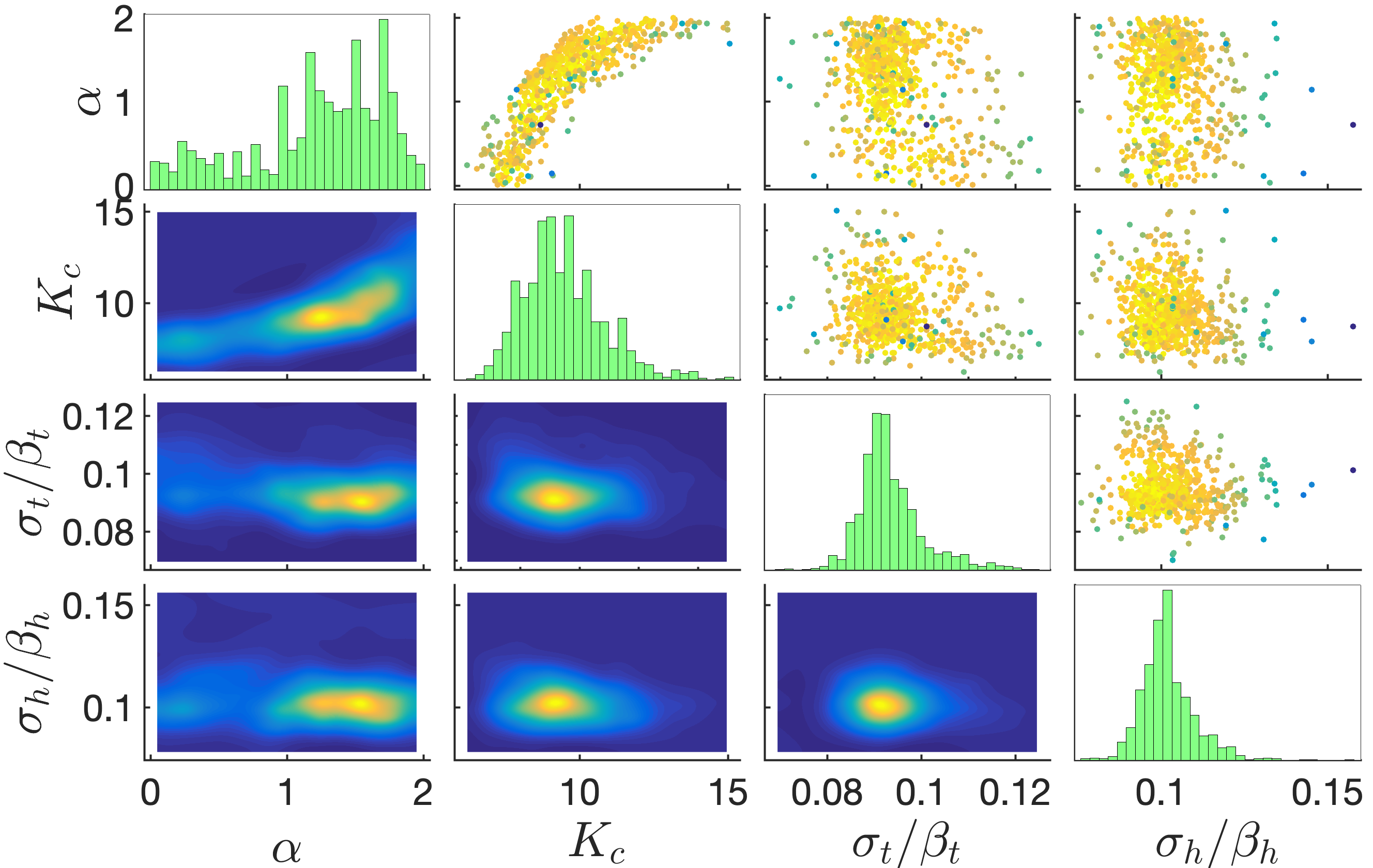}};
		\node (a) at (-6.8,-3.2) {(A) 1\% noise};
		\node (c) at (1.8,-3.2) {(B) 10\% noise};
		\end{tikzpicture}			
		\end{center}	
		\caption{Parameter estimation results for two swimmers initially 0.5 apart with fluid resistance $\alpha = 1.0$ and curvature stiffness of tail $K_C = 8.0$ using two different noise levels for the tails and the heads. In each experiment, noisy distance data between the center of the heads and tips of the tails were used as reference data. The experiment was run until $T = 20$. Histograms for each parameter are displayed along the main diagonal of the figure. Subfigures below the diagonal show the marginal joint density functions for each pair of parameters, while subfigures above the diagonal show the samples used in the final stage of TMCMC. Colors correspond to probabilities, with yellow likely and blue unlikely.}
\label{fig:two_swimmer2SD}
\end{figure}

For the 1\% noise level case, results are shown in Fig.~\ref{fig:two_swimmer2SD}(A) and Table \ref{table:two_swim_twop_two_sigma}. In the figure, a clear correlation between the fluid resistance $\alpha$ and the tail curvature stiffness $K_C$, as seen in the previous experiments. In addition, we see that the marginal distributions between the parameters and the noise appear to be uncorrelated. Finally, all nominal parameter values are recovered within two-standard deviations of the recovered means, indicating the ability of the method to recover parameters from noisy data.

 \begin{table}[h]
\centering
\begin{tabular}{c c c c c c c c c} 
 \hline
 \hline
 Noise Level & $\alpha$ & $u_{\alpha}$ (\%) & $K_C$ & $u_{K_C}$ (\%) & $\sigma_t/\beta_t$ & $u_{\sigma_t/\beta_t}$ (\%)  & $\sigma_h/\beta_h$ & $u_{\sigma_h/\beta_h}$ (\%)\\
 \hline
1\% & 1.0597 &27.27 & 1.0353 &7.56 &0.00959 & 11.18&0.0100 &10.51 \\
10\% &1.2275 &40.70 & 1.1833 & 14.22&0.09343 & 7.25&0.1020 &7.68 \\
 \hline
\end{tabular}\smallskip
\caption{Posterior means and uncertainties for parameter estimation on $\alpha$ and $K_C$ for two swimmers using distance between the heads and tails as noisy observation data and looking for two different noise levels.}
\label{table:two_swim_twop_two_sigma}
\end{table}

The 10\% noise case in Fig.~\ref{fig:two_swimmer2SD}(B) and Table \ref{table:two_swim_twop_two_sigma} shows similar results to the 1\% noise case. Again, a strong positive correlation is seen between the $\alpha$ and $K_c$ parameters and the joint marginal distributions are unimodal and ellipsoidal, indicating no correlation between the noise and the parameter values. Furthermore, all nominal parameter values are recovered within two-standard deviations of the recovered means. We note that the distributions of the parameters are more spread out than the 1\% noise case, but this is due to the added noise from the observations.

\section{Discussion and Conclusions}
Overall, the results highlight that the Bayesian framework can estimate both fluid and swimmer material parameters while quantifying uncertainty in the estimates. Since there are complex dependencies on emergent swimming speeds and trajectories, it is not surprising that the potential parameter distributions from which the data could come from grew as the noise in the data increased. Through several test cases we have highlighted that the amount of data used as well as the type of data (along with noise) must be considered carefully as it will have an effect on the ability to estimate parameters. In micro-swimmer applications, a long enough time interval must be chosen to capture the full dynamics in order to estimate the parameter with high certainty. In addition, in the case of multiple swimmers, if data is of slightly different magnitudes, using two different noise levels led to tighter distributions of parameter values. 

The ability to use this Bayesian framework to estimate parameters will aid in further understanding emergent properties of micro-swimmers since computational resources limit running models throughout the entire parameter space to understand emergent wave forms and trajectories of swimmers in different fluid environments (when using fully resolved fluid-structure interaction models). Experiments show synchronization of bull sperm beat forms \cite{Woolley09}, alignment of mouse sperm (head to tail) \cite{Moore02}, and sea urchin sperm swimming in vortices \cite{Riedel05}. Utilizing the proposed framework with noisy data of sperm trajectories, there is the potential to identify what parameter ranges or distributions that could lead to these emergent patterns. 

Although we have focused on micro-swimmers in a Brinkman fluid, we emphasize that it is possible to use this methodology in a variety of contexts. Given noisy data of an elastic micro-swimmer, parameter identification could be used to identify fluid properties e.g. visocisty in a Stokesian fluid or viscosity and/or relaxation time of elastic polymers in the fluid if viscoelastic and assuming governed by an Oldroyd B type model. Additionally, we focused on estimating the tail curvature stiffness parameter but in general, the algorithms have the potential to identify any parameters, including internal properties of swimmer, e.g. constitutive parameters (e.g. stiffness parameter or elastic moduli). 

In this paper, our chief concern was to offset the computational cost of the micro-swimmer model through minimizing the number of evaluations needed for the likelihood computation. Our TMCMC-based approach allowed us to thoroughly explore the parameter space and resulted in reliable estimates for the posterior distributions. We remark that if observed data are not fully known from the onset or if there is a prohibitively large amount of observations, filtering-based approaches \cite{stuart2010,Kaipio2005,Tarantola2005} can aid in re-sampling either to incorporate new data as they become available or to split the data for computational efficiency. 

Although we did not explore model selection, this framework can be utilized to not only estimate parameters but decide the likelihood that the data is described by a particular model. This has been done previously in the context of identifying arterial wall abnormalities \cite{Larson19b}. In the future, we could test different constitutive laws and models for micro-swimmers.

\section{Acknowledgements}
Simulations were run at the Center for Computation and Visualization at Brown University. KL and AM were partially supported by the NSF through grants DMS-1521266 and DMS-1552903. SDO was supported, in part, by NSF grant DMS-1455270.

\section*{Appendix}\label{sec:appendix}
Details of the force derivation and numerical algorithm are given here. We assume that the flagellum will beat with a  curvature wave corresponding to a low amplitude tapered  sine wave, as observed in experiments with human sperm \cite{Smith09}. This corresponds to a preferred flagellum configuration 
$$\hat{X}_F^j(s,t)=[\hat{x}(s,t),\hat{y}^j(s,t)]=[s,a^j(1-s)\sin(\eta^js-\omega^j t)],$$
where the $j$-th micro-swimmer has amplitude $a$, wavelength $2\pi/\eta$, and beat frequency $\omega/2\pi$. Here, $s$ is a parameter initialized as arc length where $0\leq s\leq 1$ (the nondimensional length of the flagellum is 1). The flagellum will attempt to reach this preferred configuration and we will define the bending energy as
\begin{equation}
E^{j}_{F,bend}=K^j_{C}\int_{\Gamma_{F}^{j}}\left(\zeta^j(s,t)-\hat{\zeta}^j(s,t)\right)^2ds,\label{EFB}
\end{equation}
where $\Gamma_F^j$ is the centerline curve corresponding to the $j$-th flagellum and $K_{C}$ is a stiffness coefficient enforcing the curvature or bending constraint. The preferred curvature $\hat{\zeta}^j$ and actual curvature $\zeta^j(s,t)$ of the flagellum are given as
\begin{equation}
    \hat{\zeta}^j=\frac{\partial^2 \hat{y}^j}{\partial s^2},\hspace{0.5cm}\zeta^j(s,t)= \frac{\frac{\partial^2 y^j}{\partial s^2}\frac{\partial x^j}{\partial s} - \frac{\partial^2 x^j}{\partial s^2}\frac{\partial y^j}{\partial s}}{\left( \left(\frac{\partial x^j}{\partial s}\right)^2 + \left(\frac{\partial y^j}{\partial s}\right)^2\right)^{3/2}},\label{curva}
\end{equation}
where $\bm{X}_F^j=[x^j,y^j]$.

In addition to the bending component, we will account for an additional energy component that will tend to maintain the inextensibility of the flagellum. This results in
\begin{equation}
    E_{F,tens}^j=\int_{\Gamma_F^j}K_{T}^j\left(\left|\left|\frac{\partial^2\bm{X}_F^j}{\partial s^2}\right|\right|-1\right)^2ds,\label{EFT}
\end{equation}
which in a discretized form, corresponds to Hookean springs between points on the flagellum with stiffness coefficient $K_{T}$.

Similar to the flagellum, we assume a preferred shape or curvature of the head. In this simple model, we will assume a head shape with radius $H_r$ and preferred curvature $\hat{\kappa}=1/H_r$. The corresponding energy is 
$$E_{H,bend}^j=\int_{\Gamma_H^j}K_{H,C}^j\left(\kappa^j(s,t)-\hat{\kappa}(s,t)\right)^2ds,$$
where $\Gamma_H^j$ corresponds to the circular head. Here, the actual curvature $\kappa^j(s,t)$ is calculated using the same equation as $\zeta$ in (\ref{curva}), but now $\bm{X}_H^j=[x^j,y^j]$. In addition, we also have an energy to maintain inextensibility in the head, the same as (\ref{EFT}) using $\Gamma_H^j$, $\bm{X}_H^j$, and $H_{C,tens}^j$, where we envision Hookean springs between points on the membrane of the head as well as springs connecting points on the circular head that are $\pi$ apart (we choose $\mathcal{N}_H$ to be even to ensure points and springs exactly $\pi$ apart).

The swimmer is initialized (left to right) to have the center of the circular head be placed with a y-coordinate the same as the rightmost point on the flagellum and an x-coordinate that is shifted to the right of the rightmost point by $H_r$  and an additional small distance apart, $dN$. To ensure that the passive head remains attached to the actively bending flagellum, and to represent the stiff neck region of a sperm, we connect the head and flagellum with 5 springs. These springs connect the rightmost point (the $\mathcal{N}_F$-th point) of the flagellum to the points on the circle with $\theta=(\pi-d\theta),\pi,(\pi+d\theta)$ where $d\theta$ is the angular spacing between the $\mathcal{N}_H$ points on the head. Additionally, there are two springs connecting the second rightmost point on the flagellum ($\mathcal{N}_F-1$) to the points on the circle with $\theta=\pi\pm d\theta$. These springs will have a stiffness coefficient $K_{N,tens}$. There is also an energy based on the desired angle between the flagellum and the head. Let $\bf{z}_1$ be the vector connecting the $\mathcal{N}_F$-th point on the flagellum and the point on the head with $\theta=\pi$ and let $\bf{z}_2$ be the vector connecting the points on the head with $\theta=\pi\pm d\theta$. In general, we wish for these vectors to be approximately orthogonal, and we can derive an energy and hence forces that penalize this deviation, tending to maintain $\bm{z}_1\cdot\bm{z}_2=0$ with stiffness coefficient $K_{N,ang}$ \cite{Fauci95}.

Given a configuration for each of the $\mathcal{M}_S$ swimmers at the initial time point, we determine the forces on the $\mathcal{M}_S\mathcal{N}_T$ discretized points using (\ref{Fenergy}), where each of the components are calculated using \eqref{EFB}--\eqref{EFT}. Second order finite difference approximations are utilized in the calculation of all derivatives in the energy components and a trapezoidal rule is used to calculate the integrals. The forces are then used to calculate the resulting velocity at points along the discretized swimmer, (\ref{BrV}). The location of the swimmer is updated using the no-slip condition, numerically implemented with a forward Euler method. The next time step is reached, where this calculation is repeated.

In these simulations, when there is more than one swimmer, we assume that the beat form parameters such as the amplitude and beat frequency are the same for each swimmer. In addition, we assume that all stiffness parameters are the same. All parameters are given in Table \ref{Tab:param}.

\begin{table}[htb!]
\centering
\caption{Parameters for swimmer model.}
\begin{tabular}{lr}\hline\hline
$L$, characteristic length scale & 100 $\mu$m \\
$\mathcal{N}_F$, points on flagellum & 50 \\
$\mathcal{N}_H$, points on head & 24 \\
$\varepsilon$, regularization parameter & 0.045\\
$K_{N,tens}$, tensile stiffness of neck & 500 \\
$K_{H,tens}$, tensile stiffness of head & 10000 \\
$K_{tens}$, tensile stiffness of flagellum & 10000 \\
$K_{C}$, curvature stiffness of flagellum & 8 \\
$K_{H,C}$, curvature stiffness of head & 0.8 \\
$K_{N,ang}$, stiffness for flagellum-body connecting angle & 10000 \\
$\triangle t$, time step & 1$\times$10$^{-4}$\\
$\eta$, wavenumber & $2\pi$ \\
$a$, amplitude & 0.1\\
$\omega$, frequency & $4\pi$ \\
\end{tabular}\label{Tab:param}
\end{table}

\bibliographystyle{unsrt}  
\bibliography{references}  

\begin{thebibliography}{10}

\bibitem{Gaffney11}
E.~A. Gaffney, H.~Gad{\^e}lha, D.~J. Smith, J.~R. Blake, and J.~C.
  Kirkman-Brown.
\newblock Mammalian sperm motility: observation and theory.
\newblock {\em Annu Rev Fluid Mech}, 43:501--528, 2011.

\bibitem{Lauga09}
E~Lauga and TR~Powers.
\newblock The hydrodynamics of swimming microorganisms.
\newblock {\em Rep Prog Phys}, 72:096601, 2009.

\bibitem{Taylor51}
GI~Taylor.
\newblock Analysis of the swimming of microscopic organisms.
\newblock {\em Proc Roy Soc Lond Ser A}, 209:447--461, 1951.

\bibitem{Taylor52}
GI~Taylor.
\newblock The action of waving cylindrical tails in propelling microscopic
  organisms.
\newblock {\em Proc Roy Soc Lond Ser A}, 211:225--239, 1952.

\bibitem{Carichino18}
L~Carichino and SD~Olson.
\newblock Emergent three-dimensional sperm motility: coupling calcium dynamics
  and preferred curvature in a kirchhoff rod model.
\newblock {\em J Math Med Biol}, 2018.
\newblock doi:10.1093/imammb/dqy015.

\bibitem{Dillon06}
RH~Dillon, LJ~Fauci, and X~Yang.
\newblock Sperm motility and multiciliary beating: an integrative mechanical
  model.
\newblock {\em Comput Math Appl}, 52:749--758, 2006.

\bibitem{Elgeti10}
J~Elgeti, UB~Kaupp, and G~Gompper.
\newblock Hydrodynamics of sperm cells near surfaces.
\newblock {\em Biophys J}, 99(4):1018--1026, 2010.

\bibitem{Elgeti15}
J~Elgeti, RG~Winkler, and G~Gompper.
\newblock Physics of microswimmers- single particle motion and collective
  behavior: a review.
\newblock {\em Rep Prog Phys}, 78:056601, 2015.

\bibitem{Huang18}
J~Huang, L~Carichino, and SD~Olson.
\newblock Hydrodynamic interactions of actuated elastic filaments near a planar
  wall with applications to sperm motility.
\newblock {\em J Coupled Syst Multiscale Dyn}, 6:163--175, 2018.

\bibitem{Ishimoto18}
K~Ishimoto and EA~Gaffney.
\newblock An elastohydrodynamical simulation study of filament and spermatozoan
  swimming driven by internal couples.
\newblock {\em IMA J Applied Math}, 83:655--679, 2018.

\bibitem{Schoeller18}
SF~Schoeller and EE~Keaveny.
\newblock Flagellar undulations to collective motion: predicting the dynamics
  of sperm suspensions.
\newblock {\em J Roy Soc Interface}, 15:20170834, 2018.

\bibitem{Smith09Surf}
DJ~Smith, EA~Gaffney, JR~Blake, and JC~Kirkman-Brown.
\newblock Human sperm accumulation near surfaces: a simulation study.
\newblock {\em J Fluid Mech}, 621:289--320, 2009.

\bibitem{Toshihiro19}
T~Omori and T~Ishikawa.
\newblock Swimming of spermatozoa in a maxwell fluid.
\newblock {\em Micromach}, 10:78, 2019.

\bibitem{Yang08}
Y~Yang, J~Elgeti, and G~Gompper.
\newblock Cooperation of sperm in two dimensions: synchronization, attraction,
  and aggregation through hydrodynamic interactions.
\newblock {\em Phys Rev E}, 78:061903--1--9, 2008.

\bibitem{Ishimoto18b}
K~Ishimoto and EA~Gaffney.
\newblock Hydrodynamic clustering of human sperm in viscoelastic fluids.
\newblock {\em Sci Rep}, 8:15600, 2018.

\bibitem{Olson11}
SD~Olson, SS~Suarez, and LJ~Fauci.
\newblock Coupling biochemistry and hydrodynamics captures hyperactivated sperm
  motility in a simple flagellar model.
\newblock {\em J Theor Bio}, 283:203--216, 2011.

\bibitem{Teran10}
J~Teran, L~Fauci, and M~Shelley.
\newblock Viscoelastic fluid response can increase the speed of a free swimmer.
\newblock {\em Phys Rev Lett}, 104:038101--4, 2010.

\bibitem{Thomases14}
B~Thomases and RD~Guy.
\newblock Mechanisms of elastic enhancement and hindrance for finite-length
  undulatory swimmers in viscoelastic fluids.
\newblock {\em Phys Rev Lett}, 113:098102, Aug 2014.

\bibitem{Mortimer00}
ST~Mortimer.
\newblock {CASA}—practical aspects.
\newblock {\em J Androl}, 21:515--524, 2000.

\bibitem{Dominic09}
DW~Dominic, W~Pelle, CJ~Brokaw, KA~Lesich, and Lindemann CB.
\newblock Mechanical properties of the passive sea urchin sperm flagellum.
\newblock {\em Cell Motil Cytoskel}, 66:721--735, 2009.

\bibitem{Gadelha13}
H~Gadelha, EA~Gaffney, and A~Goriely.
\newblock The counterbend phenomenon in flagellar axonemes and cross-linked
  filament bundles.
\newblock {\em Proc Natl Acad Sci USA}, 110:12180--12195, 2013.

\bibitem{Riedel07}
IH~Riedel-Kruse, A~Hilfinger, J~Howard, and F~Julicher.
\newblock How molecular motors shape the flagellar beat.
\newblock {\em HFSP J}, 1:192--208, 2007.

\bibitem{Gao14}
W~Gao and J~Wang.
\newblock Synthetic micro/nanomotors in drug delivery.
\newblock {\em Nanoscale}, 6:10486--10494, 2014.

\bibitem{Nelson10}
BJ1 Nelson, IK~Kaliakatsos, and JJ~Abbott.
\newblock Microrobots for minimally invasive medicine.
\newblock {\em Annu Rev Biomed Eng}, 12:55--85, 2010.

\bibitem{Sanders09}
L~Sanders.
\newblock Microswimmers make a splash: tiny travelers take on a viscous world.
\newblock {\em Science News}, 176:22--25, 2009.

\bibitem{Tierno08}
P~Tierno, R~Golestanian, I~Pagonabarraga, and F~Sagues.
\newblock Magnetically actuated colloidal micro swimmers.
\newblock {\em J Phys Chem B}, 112:16525--16528, 2008.

\bibitem{Plouraboue17}
F~Plouraboue, Thiam EI, B~Delmotte, and E~Climent.
\newblock Identification of internal properties of fibres and micro-swimmers.
\newblock {\em Proc Roy Soc A}, 473:20160517, 2017.

\bibitem{Tsang19}
ACH Tsang, PW~Tong, Nallan S, and OS~Pak.
\newblock Self-learning how to swim at low {R}eynolds number.
\newblock 2019.

\bibitem{Cortez10}
R~Cortez, B~Cummins, K~Leiderman, and D~Varela.
\newblock Computation of three-dimensional {B}rinkman flows using regularized
  methods.
\newblock {\em J Comput Phys}, 229:7609--7624, 2010.

\bibitem{Leiderman16}
K~Leiderman and SD~Olson.
\newblock Swimming in a two-dimensional brinkman fluid: Computational modeling
  and regularized solutions.
\newblock {\em Phys Fluids}, 28(2):021902, 2016.

\bibitem{Ho19}
N.~Ho, K.~Leiderman, and S.D. Olson.
\newblock A 3-dimensional model of flagellar swimming in a {B}rinkman fluid.
\newblock {\em J Fluid Mech}, 864:1088--1124, 2019.

\bibitem{Hadjidoukas15}
P.E. Hadjidoukas, P.~Angelikopoulos, C.~Papadimitriou, and P.~Koumoutsakos.
\newblock {$\Pi$}4{U}: A high performance computing framework for bayesian
  uncertainty quantification of complex models.
\newblock {\em J Comp Phys}, 284:1 -- 21, 2015.

\bibitem{Bowman18}
Clark Bowman, Karen Larson, Alexander Roitershtein, Derek Stein, and Anastasios
  Matzavinos.
\newblock {\em Bayesian Uncertainty Quantification for Particle-Based
  Simulation of Lipid Bilayer Membranes}, pages 77--102.
\newblock Springer International Publishing, Cham, 2018.

\bibitem{Larson19}
Karen Larson, Loukas Zagkos, Mark~Mc Auley, Jason Roberts, Nikos~I. Kavallaris,
  and Anastasios Matzavinos.
\newblock Data-driven selection and parameter estimation for {DNA} methylation
  mathematical models.
\newblock {\em J Theor Biol}, 467:87 -- 99, 2019.

\bibitem{Rutllant05}
J.~Rutllant, M.~Lopez-Bejar, and F.~Lopez-Gatius.
\newblock Ultrastructural and rheological properties of bovine vaginal fluid
  and its relation to sperm motility and fertilization: a review.
\newblock {\em Reprod Dom Anim}, 40:79--86, 2005.

\bibitem{Saltzman94}
W.~M. Saltzman, M.~L. Radomsky, K.~J. Whaley, and R.~A. Cone.
\newblock Antibody diffusion in human cervical mucus.
\newblock {\em Biophys J}, 66:508, 1994.

\bibitem{Flemming10}
HC~Flemming and Wingender J.
\newblock The biofilm matrix.
\newblock {\em Nat Rev Microbiol}, 8:623--633, 2010.

\bibitem{Miradbagheri16}
SA~Miradbagheri and HC~Fu.
\newblock Helicobacter pylori couples motility and diffusion to actively create
  a heterogeneous complex medium in gastric diseases.
\newblock {\em Phys Rev Lett}, 116, 2016.

\bibitem{Auriault09}
J.~L. Auriault.
\newblock On the domain of validity of {B}rinkman's equation.
\newblock {\em Trans. Porous Media}, 79:215--223, 2009.

\bibitem{Brinkman47}
H.~C. Brinkman.
\newblock A calculation of the viscous force exerted by a flowing fluid on a
  dense swarm of paticles.
\newblock {\em Appl Sci Res}, 1:27--34, 1947.

\bibitem{Durlofsky87}
L.~Durlofsky and J.~F. Brady.
\newblock Analysis of the {B}rinkman equation as a model for flow in porous
  media.
\newblock {\em Phys Fluids}, 30(11):3329--3341, 1987.

\bibitem{Howells74}
I.~D. Howells.
\newblock Drag due to the motion of a {N}ewtonian fluid through a sparse random
  array of small fixed rigid objects.
\newblock {\em J Fluid Mech}, 64:449--475, 1974.

\bibitem{Spielman68}
L.~Spielman and S.~L. Goren.
\newblock Model for predicting pressure drop and filtration efficiency in
  fibrous media.
\newblock {\em Env Science Tech}, 1(4):279--287, 1968.

\bibitem{Fu10}
H~Fu, VB~Shenoy, and TR~Powers.
\newblock Low {R}eynolds number swimming in gels.
\newblock {\em Europhys Lett}, 91, 2010.

\bibitem{Leshansky09}
AM~Leshansky.
\newblock Enhanced low-{R}eynolds-number propulsion in heterogenous viscous
  environments.
\newblock {\em Phys Rev E}, 80:051911, 2009.

\bibitem{Morandotti12}
M~Morandotti.
\newblock Self-propelled micro-swimmers in a {B}rinkman fluid.
\newblock {\em J Biol Dynamics}, 6:88--103, 2012.

\bibitem{Nganguia18}
H~Nganguia and OS~Pak.
\newblock Squirming motion in a {B}rinkman medium.
\newblock {\em J Fluid Mech}, 855:554--573, 2018.

\bibitem{Ho16}
N~Ho, K~Leiderman, and SD~Olson.
\newblock Swimming speeds of filaments in viscous fluids with resistance.
\newblock {\em Phys Rev E}, 93(4):043108, 2016.

\bibitem{Olson15}
SD~Olson and K~Leiderman.
\newblock Effect of fluid resistance on symmetric and asymmetric flagellar
  waveforms.
\newblock {\em J Aero Aqua Bio-mech}, 4(1):12--17, 2015.

\bibitem{Peskin02}
C.S. Peskin.
\newblock The immersed boundary method.
\newblock {\em Acta Numer}, 11:459--517, 2002.

\bibitem{Leiderman17}
K.~Leiderman and S.D. Olson.
\newblock Erratum: “swimming in a two-dimensional brinkman fluid:
  Computational modeling and regularized solutions” [phys. fluids 28, 021902
  (2016)].
\newblock {\em Phys Fluids}, 29:029901, 2017.

\bibitem{Ahmadi17}
E~Ahmadi, R~Cortez, and H~Fujioka.
\newblock Boundary integral formulation for flows containing an interface
  between two porous media.
\newblock {\em J Fluid Mech}, 816:71--93, 2017.

\bibitem{Fauci95}
LJ~Fauci and A~McDonald.
\newblock Sperm motility in the presence of boundaries.
\newblock {\em Bull Math Biol}, 57:679--699, 1995.

\bibitem{Beck04}
James~L. Beck and Ka-Veng Yuen.
\newblock Model selection using response measurements: Bayesian probabilistic
  approach.
\newblock {\em J Eng Mech}, 130(2):192--203, 2004.

\bibitem{Vanik00}
M.~W. Vanik, J.~L. Beck, and S.~K. Au.
\newblock Bayesian probabilistic approach to structural health monitoring.
\newblock {\em J Eng Mech}, 126(7):738--745, 2000.

\bibitem{Ching07}
J.Y. Ching and Y.C. Chen.
\newblock Transitional markov chain monte carlo method for bayesian model
  updating, model class selection, and model averaging.
\newblock {\em J Eng Mech}, 133:816--832, 2007.

\bibitem{Woolley09}
DM~Woolley, RF~Crockett, WDI Groom, and SG~Revell.
\newblock A study of synchronisation between the flagella of bull spermatozoa,
  with related observations.
\newblock {\em J Exp Biol}, 212:2215--2223, 2009.

\bibitem{Moore02}
H~Moore, K~Dvorakova, N~Jenkins, and W~Breed.
\newblock Exceptional sperm cooperation in the wood mouse.
\newblock {\em Nature}, 418:174--177, 2002.

\bibitem{Riedel05}
IH~Riedel, K~Kruse, and J~Howard.
\newblock A self-organized vortex array of hydrodynamically entrained sperm
  cells.
\newblock {\em Science}, 309:300--303, 2005.

\bibitem{stuart2010}
AM~Stuart.
\newblock Inverse problems: A {B}ayesian perspective.
\newblock {\em Acta Numerica}, 19:451--559, 2010.

\bibitem{Kaipio2005}
J~Kaipio and E~Somersalo.
\newblock {\em Statistical and Computational Inverse Problems}, volume 160.
\newblock Springer, 2005.

\bibitem{Tarantola2005}
Albert Tarantola.
\newblock {\em Inverse Problem Theory and Methods for Model Parameter
  Estimation}.
\newblock SIAM, 2005.

\bibitem{Larson19b}
K~Larson, C~Bowman, C~Papadimitriou, P~Koumoutsakos, and A~Matzavinos.
\newblock Detection of arterial wall abnormalities via {B}ayesian model
  selection.
\newblock {\em Royal Society Open Science}, 6:182229, 2019.

\bibitem{Smith09}
DJ~Smith, EA~Gaffney, H~Gad{\^e}lha, N~Kapur, and JC~Kirkman-Brown.
\newblock Bend propagation in the flagella of migrating human sperm, and its
  modulation by viscosity.
\newblock {\em Cell Motil Cytoskel}, 66(4):220--236, 2009.

\end{thebibliography}

\end{document}